Effects of Yttrium doping on Proton Conductivity in Ba (Fe, Co, Zr, Y) $O_{3-\delta}$ Cathode Materials for Proton Ceramic Fuel Cells

Chiyoung Kim[a], Ryan Jacobs[a], Jun Meng[a], Jack H. Duffy[b], Kyle S. Brinkman[b], Yueh-Lin Lee[c], Harry W. Abernathy[c], Dane Morgan[a]

a. Department of Materials Science and Engineering, University of Wisconsin – Madison, Madison, Wisconsin 53706, United States
b. Department of Materials Science and Engineering, Clemson University, Clemson, South Carolina 29634, United States
c. National Energy Technology Laboratory, United States Department of Energy, Morgantown, West Virginia 26507, United States

## Abstract

Protonic ceramic fuel cells (PCFCs) offer efficient energy conversion at intermediate temperatures, but their performance is often constrained by sluggish oxygen reduction kinetics at the cathode. Triple conducting perovskite materials such as $BaCo_xFe_yZr_zY_{1-x-y-z}O_{3-\delta}$ (BCFZY) have shown promise as air electrode candidates due to their ability to conduct electrons, oxygen ions, and protons simultaneously. Among design strategies to enhance proton conduction, Yttrium (Y) doping is widely adopted, though its specific role in governing proton transport remains unclear. In this study, we investigated the impact of Y doping on proton conductivity in the BCFZY system by decoupling the effects on proton diffusivity and proton concentration. Using *ab initio* molecular dynamics (AIMD) simulations, we assessed the proton tracer diffusivity across three compositions: $BaCo_{0.4}Fe_{0.4}Zr_{0.2}O_{3-\delta}$ (BCFZ), $BaCo_{0.4}Fe_{0.4}Zr_{0.1}Y_{0.1}O_{3-\delta}$ (BCFZY0.1) and $BaCo_{0.4}Fe_{0.4}Y_{0.2}O_{3-\delta}$ (BCFY). Thermodynamic defect models incorporating hydration, oxidation, and hydride formation reactions are employed to estimate temperature-dependent proton concentrations under PCFC operating gas conditions and low oxygen partial pressure ($pO_2$). Our results suggest that increasing the Y content in the BCFZY system from BCFZ to BCFY slightly increases proton diffusivity by 1.7 % at 500 °C. Under PCFC operating gas conditions, we predict that proton concentration and proton conductivity increase by 3.8 % and 5.3 %, respectively, at 500 °C. However, these increases under PCFC operating gas conditions are relatively modest compared with the larger enhancements in proton conductivity reported experimentally under reducing conditions, highlighting the importance of gas conditions in evaluating the effects of Y doping. Under these low oxygen partial pressure conditions, we predict that the dominant proton transport mechanism shifts from hydroxyl groups ($OH_O^{\cdot}$) to hydride species ($H_O^{\cdot}$). This work provides mechanistic insight into proton states, including hydroxyl species ($OH_O^{\cdot}$) and hydride species ($H_O^{\cdot}$), as well as proton transport in BCFZY cathodes, and offers design principles for optimizing proton-conducting air electrodes in next-generation PCFCs.

## 1. Introduction

Solid oxide fuel cells (SOFCs) are highly efficient energy conversion devices that operate on a range of fuels, including hydrogen and hydrocarbons. [1,2] However, their requirement for elevated operating temperatures (typically around 800 °C) leads to significant challenges, such as increased material degradation and system costs. [3] Protonic ceramic fuel cells (PCFCs) were introduced as a promising alternative, offering comparable electrochemical efficiency at substantially reduced temperatures (400–650 °C). [4] This improvement is attributed to the superior mobility of protons relative to oxygen ions in the electrolyte. [5] Nonetheless, under these intermediate temperatures, the performance of PCFCs remains constrained, primarily by sluggish cathodic oxygen reduction reaction (ORR) kinetics. [6,7] As a result, the development of improved cathode materials remains a critical need, particularly those capable of facilitating the transport of all three charge carriers: protons, oxygen ions, and electrons. [8] Such triple-conducting materials can expand the electrochemically active area and accelerate overall reaction kinetics. [9]

In this context, the $BaCo_xFe_yZr_zY_{1-x-y-z}O_{3-\delta}$ (BCFZY) perovskite family has emerged as a benchmark class of materials for PCFC cathodes due to their ability to simultaneously conduct protons, electrons, and oxygen ions. [5,10] BCFZY materials are derived from strategic modifications of the $BaZrO_3$ perovskite, initially doped with redox-active Fe and Co to create $BaCo_{0.4}Fe_{0.4}Zr_{0.2}O_{3-\delta}$ (BCFZ), and further doped with $Y^{3+}$ to form compositions such as $BaCo_{0.4}Fe_{0.4}Zr_{0.1}Y_{0.1}O_{3-\delta}$ (BCFZY0.1) and $BaCo_{0.4}Fe_{0.4}Y_{0.2}O_{3-\delta}$ (BCFY). [11,12] Yttrium doping of perovskite oxides has been reported to improve proton uptake by generating oxygen vacancies and localizing electrons on oxygen ions. [12] Relative to undoped BCFZ, Y-doped BCFZY0.1 has been reported to exhibit significant changes in area-specific resistance (ASR), proton conductivity, and oxygen conductivity, with an order-of-magnitude lower ASR measured in air at 500 °C, up to sixfold higher proton conductivity measured under reducing (low-$pO_2$) conditions, and a threefold decrease in oxygen conductivity measured in air at 550 °C. [5,13,14]

However, despite these experimental observations, the mechanistic role of Y doping in determining proton transport properties, including both proton diffusion coefficients and proton concentrations, remains poorly understood.[13] Additionally, the relationship between proton and oxygen conductivity and their respective contributions to electrode performance (e.g., ASR) is also not yet well understood. These gaps in knowledge limit our ability to design cathodes that fully exploit the benefits of Y substitution and achieve optimized triple conducting behavior.

Furthermore, although experimental studies have demonstrated the presence of proton conduction in BCFZY systems, these measurements are often sensitive to environmental conditions (e.g., humidity, oxygen partial pressure), complicating the interpretation of underlying transport mechanisms.[13,15,16] Computational studies, particularly *ab initio* molecular dynamics (AIMD), offer a valuable tool to probe proton transport at the atomic scale while controlling for such external factors.[17] To date, however, no AIMD studies have directly evaluated proton migration behavior in BCFZY systems across varying Y concentrations, nor have they systematically isolated the effects of Y doping on proton mobility versus proton incorporation. In our prior investigation of oxygen conductivity in BCFZY [18], we employed ab initio molecular dynamics (AIMD) simulations combined with thermodynamic defect modelling to examine oxygen conductivity trends in BCFZY materials. The findings suggest that Y doping leads to a modest reduction in oxygen conductivity, decreasing from 437 ± 136 mS/cm in BCFZ to 254 ± 102 mS/cm in BCFZY0.1 at 500 °C. The corresponding activation energies were 0.15 ± 0.01 eV and 0.17 ± 0.01 eV, respectively. With further Y doping, BCFY exhibited a lower oxygen conductivity (121 ± 49 mS/cm) and a higher activation energy of 0.26 ± 0.01 eV, indicating a progressive reduction in oxygen conductivity with increasing Y content. These results, consistent with previous experimental studies, suggest that the benefits of Y come primarily from its impact on proton conductivity.

In this study, we investigated the influence of Y doping on proton conductivity in BCFZY cathode materials using a combination of *ab initio* molecular dynamics and thermodynamic defect modeling. We examined three compositions: $BaCo_{0.4}Fe_{0.4}Zr_{0.2}O_{3-\delta}$ (BCFZ), $Ba\,Co_{0.4}Fe_{0.4}Zr_{0.1}Y_{0.1}O_{3-\delta}$ (BCFZY0.1), and $BaCo_{0.4}Fe_{0.4}Y_{0.2}O_{3-\delta}$ (BCFY). AIMD simulations were used to determine proton tracer diffusivities and their associated activation energies, while thermodynamic models incorporating hydration, oxidation, and hydride formation reactions are used to calculate temperature-dependent proton concentrations.

A central goal of this study is to understand how Y doping influences the underlying factors that determine proton conductivity in BCFZY, specifically proton diffusivity ($D_H$) and proton concentration ($c_H$) under PCFC operating gas conditions. Our results suggest that Y doping in the BCFZY system has a minor impact on proton diffusivity calculated using AIMD and proton concentration under PCFC operating gas conditions, and therefore a minor impact on the resulting proton conductivity calculated using the Nernst-Einstein relation. We predict that the dominant proton transport mechanism shifts from hydroxyl groups ($OH_O^{\cdot}$) to hydride species ($H_O^{\cdot}$) under reducing, low oxygen partial pressure conditions. By decoupling and quantifying the contributions of proton diffusivity ($D_H$) and proton concentration ($c_H$) to total proton conductivity, this work provides mechanistic insights into the role of acceptor dopants such as Y. It further elucidates proton states, including hydroxyl ($OH_O^{\cdot}$) groups and hydride ($H_O^{\cdot}$) species, and their transport behavior in BCFZY cathodes, while establishing a framework for evaluating dopant strategies and offering design principles for optimizing proton-conducting air electrodes in next-generation PCFCs.

## 2. Experimental section

The essential input and output files from the calculations described below are all available digitally. This includes the specific atomic coordinates used for all the ab initio calculations. Please see the Data Availability Statement.

### 2.1 Ab Initio Molecular Dynamics (AIMD) Simulations for Proton Diffusivity

To investigate proton diffusion in BCFZY systems, we performed density functional theory (DFT) simulations using the Vienna Ab initio Simulation Package (VASP).[19] The exchange-correlation energy was treated within the generalized gradient approximation (GGA) using the Perdew–Burke–Ernzerhof (PBE) functional, with a Hubbard U correction ($U_Fe$ = 5.3 eV and $U_Co$ = 3.32 eV) applied to better capture the localized d-electron behavior of transition metal sites.[20,21] Core electrons were described using projector augmented-wave (PAW) pseudopotentials and the plane-wave energy cutoff was set to 500 eV. [22]

Each simulation employed a 2×2×2 supercell derived from the 5-atom $BaZrO_3$ perovskite unit cell, resulting in a 40-atom configuration. The stoichiometries for BCFZ, BCFZY0.1, and BCFY consisted of 8 Ba, 3 Fe, 3 Co, and 24 oxygen atoms, along with 1 hydrogen atom, where Fe and Co account for 37.5 % of the B-site cations due to supercell size limitations. In constructing supercells for BCFZ, BCFZY, and BCFY compositions, the substitution of Zr and Y was carried out systematically. BCFZ contained two Zr atoms, BCFZY included one Zr and one Y atom, and BCFY contained two Y atoms.

For each of the BCFZ, BCFZY0.1 and BCFY compositions, the structures were initially relaxed without symmetry constraints to obtain their lowest-energy configurations. The resulting volumes were then used to construct pseudo cubic cells by setting equal lattice parameters (a = b = c) while maintaining the total volume. Within these fixed-volume pseudo cubic cells, a second relaxation was performed, this time only allowing ionic positions to adjust. A 4×4×4 Monkhorst-Pack k-point mesh was used for Brillouin zone sampling during relaxation of the 2×2×2 supercell (40 atoms), based on the cubic perovskite primitive unit cell (5 atoms). [23]

The oxygen non-stoichiometries (δ) in the computational cells for BCFZY systems were determined so as to achieve as close as possible to the following assigned valence states to the redox-active elements. Although Fe and Co can exhibit variable valence states they were assumed to be in the 3+ state, which is approximately consistent with their behavior under experimental conditions. [24,25] Ba, Zr, Y, and O are expected to be redox-inactive species and were treated with fixed oxidation states of 2+, 4+, 3+, and 2−, respectively. Hydrogen was included in the supercells as either a 1+ or 1- charged species to simulate hydroxyl ($OH_O^{\cdot}$) and hydride ($H_O^{\cdot}$) incorporation, respectively. O was treated as 2- and vacancies as neutral. Based on these values, the number of oxygen vacancies in a supercell was set to make the overall formal valence of the cell as close to zero as possible, and any deviation from zero was removed by adding or subtracting electrons from the cell (a charged supercell calculation). The number of electrons was manually adjusted to preserve the assigned oxidation states of Ba as 2+, Zr as 4+, Y as 3+, Fe and Co as 3+, and O as 2−. This procedure resulted in two oxygen vacancies in BCFZ, three in BCFZY0.1 and four in BCFY, corresponding to oxygen non-stoichiometries of 0.25 for BCFZ, 0.375 for BCFZY0.1 and 0.5 for BCFY, respectively.

Following structural optimization, AIMD simulations were carried out under an NVT ensemble using a Nose–Hoover thermostat at four different temperatures: 1000 K, 1333 K, 1666 K, and 2000 K, with a time step of 1 fs for a total simulation time of 100 ps.[26,27] A gamma-point sampling of reciprocal space was employed. The effect of thermal expansion on proton tracer diffusivity was investigated and found to be negligible (Figure S1).

As a practical approximation to the high-temperature paramagnetic state encountered under operating conditions, a ferromagnetic configuration was employed.[28] The simulations were initialized with a ferromagnetic ordering in which all $Fe^{3+}$ and $Co^{3+}$ ions were assigned high-spin configurations with aligned spins. Magnetic moments were set to 5 μB for $Fe^{3+}$ and 4 μB for $Co^{3+}$, reflecting their $d^5$ and $d^6$ electron configurations, respectively. The net magnetic moment of the 2×2×2 supercell was constrained to 27 μB throughout the initial structural relaxations and AIMD runs, while allowing the internal spin configuration to evolve dynamically. This approach serves as a practical approximation of magnetic disorder at elevated temperatures, while keeping the simulations computationally efficient.

To investigate the influence of oxygen non-stoichiometry on proton diffusivity in BCFZ, a series of ab initio molecular dynamics (AIMD) simulations were conducted using supercells with varying numbers of oxygen vacancies. Each simulation was run for a total of 100 ps. The supercell composition was fixed as $Ba_8Co_3Fe_3Zr_2H_1O_{(24-n)}$, where *n* represents the number of oxygen vacancies, ranging from 0 to 3. This allowed for a systematic evaluation of the impact of δ on proton diffusion behavior, providing insight into how vacancy concentration modulates proton mobility in the perovskite lattice. (Figure S2)

To obtain a representative solid solution diffusivity, AIMD simulations were performed across six different B-site cation configurations. These configurations were constructed by systematically varying the distances between Zr and Y atoms on the B-sites, covering short, medium, and long separations for Zr–Zr, Zr–Y, and Y–Y pairs. For each dopant arrangement, two distinct random distributions of Fe and Co atoms were generated. The resulting six configurations were averaged to account for the variability in local atomic environments and provide a reliable estimate of the bulk proton diffusivity in the BCFZY system. (Figure S3)

## 2.2 Proton Diffusivity Calculations

To evaluate proton self-diffusivity in the BCFZY systems, the time-averaged mean square displacement (TAMSD) method was employed.[29,30] This approach enables robust statistical averaging over trajectory segments to extract reliable diffusion coefficients from molecular dynamics simulations. The standard error of proton diffusivity was determined based on the variability across six distinct B-site configurations. The associated uncertainty in the proton migration energy ($E_{mig}$) was derived by propagating these errors through an Arrhenius analysis of diffusivity as a function of temperature, employing a weighted least-squares fitting method that incorporated the standard error values.

## 2.3 Series model fitting of proton diffusivity

$$D_{total} = \frac{(L+d)}{\frac{L}{D_b}+\frac{d}{D_{gb}}} \quad (1)$$

(L: grain size, d: grain boundary thickness)

To incorporate grain boundary effects on proton diffusivity, we made use of a series model for the total diffusion coefficient ($D_{total}$) as a function of the bulk ($D_b$) and grain boundary ($D_{gb}$) diffusion coefficients. We assume all these $D$ values are well approximated by tracer diffusion coefficients. The relationship is given in Eq. (1), [31] where $L$ is the grain size and $d$ is the grain boundary thickness. We fit $D_{gb}$ by assuming that $D_b$ is the proton diffusivity obtained from our AIMD calculations and $D_{total}$ is the experimental proton diffusivity. Among the BCFZY systems, experimental proton diffusivity data were only available for BCFZY, and therefore these data were used for fitting. [24,32] For references reporting proton tracer diffusivity, isotopic hydrogen was used during measurement.[24] Accordingly, the isotope effect was removed from the reported tracer diffusivity by applying a factor of $\sqrt{2}$. [33] For references reporting proton chemical diffusivity [32], we assumed sufficiently dilute proton concentrations such that the chemical diffusivity approximates the tracer diffusivity. For the fitting process, the bulk proton diffusivity was initially calculated using Equation (2). The terms $x_{OH}$ and $x_{Ho}$ represent the fractional concentrations of hydroxyl groups ($[OH_O^{\cdot}]$) and hydride species ($[H_O^{\cdot}]$), respectively, as determined in Section 2.5. $D_{OH}$ is the tracer diffusion of $OH_O^{\cdot}$ and was calculated from AIMD simulations of proton diffusion in BCFZY (Figure 1), while $D_{\mathrm{Ho}}$ is the tracer diffusion of $H_O^{\cdot}$ and was obtained from AIMD simulations of oxygen vacancy diffusion (Figure S7), as $H_O^{\cdot}$ is assumed to migrate via oxygen vacancy hopping.

$$D_b = x_{OH}D_{OH} + x_{Ho}D_{Ho} \quad (2)$$

$$(x_{OH} + x_{Ho} = 1)$$

To simplify the fitting procedure, we assumed an Arrhenius expression for $D_{gb}$. The prefactor ($A_{gb}$) and activation energy ($E_{gb}$) were determined such that the total diffusivity ($D_{total}$) achieves the best least-squares fit to the experimental proton diffusivity data. The resulting grain boundary Arrhenius parameters obtained for BCFZY were subsequently applied to the BCFZ and BCFY systems. Using the bulk proton diffusivity of BCFZ and BCFY from AIMD calculations, we then calculated the corresponding total diffusivities ($D_{total}$) for these systems.

## 2.4 Size Effect Analysis via Universal Interatomic Potentials

To assess the influence of simulation cell size on proton diffusion in BCFZY systems, molecular dynamics (MD) simulations were conducted using the Universal Machine Learning Interatomic Potential (U-MLIP) based on the 3-body Materials Graph Network (M3GNet).[34] These simulations aimed to reduce computational cost while maintaining fidelity with DFT-based results. The calculations were performed using the Atomic Simulation Environment (ASE) Python library under an NVT ensemble. Simulations were run for up to 100 picoseconds with a 1 femtosecond time step for five supercell sizes: 2×2×2 (40 atoms), 4×4×4 (320 atoms), 8×8×8 (2560 atoms), 12×12×12 (8640 atoms) and 16×16×16 (20480 atoms). The 2×2×2 supercell contained 8 Ba, 3 Co, 3 Fe, 2 Zr, 22 O, and 1 H atom. The proton diffusion pathways and activation energies obtained from M3GNet-based MD simulations were found to be consistent with those calculated using PBE+U AIMD methods, validating the use of M3GNet for studying proton transport in BCFZY materials. (Figure S8) These results suggest that system size does not have a systematic effect on proton diffusion under the simulated conditions.

## 2.5 Thermodynamic Defect Modeling for Proton Concentration

Proton concentrations (hydroxyl groups: $[OH_O^{\cdot}]$, hydride species: $[H_O^{\cdot}]$) were estimated using a thermodynamic defect model incorporating four key reactions: hydration, oxidation, hydride formation, and charge disproportionation.[35,36] Our modeling efforts drew upon a framework developed in the literature for BCFZ and BCFZY0.1 [24,25,37,38], but made multiple additions. The largest change is that the model was extended with additional defect reactions, namely hydride formation and charge disproportionation. We further refined the model for the oxidation thermodynamics by parameterizing the standard enthalpy of oxidation, $\Delta H^{\circ}_{Ox}$, as a function of both hole concentration and temperature, rather than using a constant value derived from a van't Hoff plot. The standard entropy of oxidation, $\Delta S^{\circ}_{Ox}$(T), was estimated by combining the solid-state vibrational entropy calculated using the Einstein model with the gas-phase O2 entropy obtained from the Shomate equation. Both the enthalpy and entropy of oxidation are described in detail in this section. These modifications enable the model to operate over a broader range of gas partial pressures. Furthermore, incorporating $Fe^{2+}$ into the defect chemistry yields a more realistic depiction of hole transport via the charge disproportionation pathway. To adapt the model to relevant gas environments, we applied gas conditions corresponding to PCFC operating conditions, as well as low oxygen partial pressure conditions derived from the thermodynamic environment of experiments used for comparison (detailed in Section 2.5).

For the mass action equation of hydration, oxidation and hydride formation reaction applied to BCFZY0.1 and BCFY, it was assumed that Y doping influences only the charge neutrality condition. Based on this, the original reactions formulated for BCFZ were revised using the charge neutrality expression:

$$[Fe_B^{\cdot}] + 2[V_O^{\cdot\cdot}] + q[M_B^q] + [OH_O^{\cdot}] + [H_O^{\cdot}] - [Fe_B'] - [Ba_A'] = 0 \quad (3)$$

Here, [X] denotes the number of species X per $ABO_3$ formula unit. Hydride species ($H_O^{\cdot}$) occupy oxygen lattice sites without contributing oxygen atoms. Therefore, the total oxygen deficiency is given by (δ + $[H_O^{\cdot}]$), where δ represents the concentration of empty oxygen vacancies ($[V_O^{\cdot\cdot}]$). Note that this differs from the usual case where δ represents the total oxygen deficiency. The host lattice is defined by the ionic framework $A^{3+}B^{3+}O_3^{2-}$. The parameter q denotes the effective charge of B-site dopants relative to $Fe^{3+}$. For BCFZ, the term $q[M_B^q]$ (corresponding to $[Zr_B^{\cdot}]$) is 0.2, while for BCFZY0.1 it is 0.1, and for BCFY, it becomes 0. The link between iron valence states and oxygen vacancy concentrations is thus governed by the level of Y substitution, as $[Ba_A']$ remains fixed at 1. Variations in Fe valence due to Y doping modify the defect chemistry, which in turn affects the predicted concentrations of charged defects.

The hydration reaction can be written as:

$$H_2O + V_O^{\cdot\cdot} + O_O^{\times} \rightleftharpoons 2OH_O^{\cdot} \quad (4)$$

The equilibrium constant for this hydration reaction, $K_{hydrat}$, is expressed as:

$$K_{hydrat} = \frac{[OH_O^{\cdot}]^2}{pH_2O\delta(3-\delta-[OH_O^{\cdot}]-[H_O^{\cdot}])} = exp^{-(\Delta H^{\circ}_{hydrat}-T\Delta S^{\circ}_{hydrat})/RT} \quad (5)$$

In the hydration reaction, water vapor ($H_2O$) is incorporated into oxygen vacancies ($V_O^{\cdot\cdot}$), forming hydroxyl groups ($OH_O^{\cdot}$). This process increases the proton concentration in the material and is a well-established mechanism for proton uptake in perovskite oxides.[12] For the standard hydration enthalpy ($\Delta H^{\circ}_{hydrat}$) and entropy ($\Delta S^{\circ}_{hydrat}$), we adopted values reported for BCFZY by Shin et al. [24] specifically $\Delta H^{\circ}_{hydrat}$ = −28.3 kJ/mol and $\Delta S^{\circ}_{hydrat}$ = −0.116 kJ/mol·K.

$$\frac{1}{2} O_2 + V_O^{\cdot\cdot} + 2Fe_{Fe}^{\times} \rightleftharpoons O_O^{\times} + 2Fe_{Fe}^{\cdot} \quad (6)$$

Equation 6 represents the oxidation reaction, where molecular oxygen ($O_2$) is incorporated into oxygen vacancies, accompanied by the generation of two holes ($h^{\cdot}$). These holes are assumed to localize on iron atoms, forming $Fe^{4+}$ species ($Fe_{Fe}^{\cdot}$).[25] Thus, iron acts as the primary hole-trapping center in the system, and charge transport occurs through these localized defect states. We note that this assumption probably warrants further study and is not essential to the model. If holes reduce Co the model would be identical for the systems studied here since we have the same concentration of Fe and Co in our systems. Even if holes are shared between the two species it might only create a modest perturbation to the model due to changes in entropy terms. However, partial or total delocalization of the holes could change the model significantly.

$$K_{OX} = \frac{(3-\delta-[OH_O^{\cdot}]-[H_O^{\cdot}])[Fe_{Fe}^{\cdot}]^2}{\sqrt{pO_2}\delta[Fe_{Fe}^{\times}]^2} = exp^{-(\Delta H^{\circ}_{Ox}-T\Delta S^{\circ}_{Ox})/RT} \quad (7)$$

The oxidation enthalpy $\Delta H^{\circ}_{Ox}$ was assumed to depend on both hole concentration and temperature and was derived by fitting oxygen non-stoichiometry (δ) to experimental data under varying oxygen partial pressure conditions for BCFZ (Zohourian et al.[25]) and BCFZY (Shin et al.[24]). The dependence on hole concentration is assumed to be due to hole-hole interactions and represented by the term $\Delta H^{\circ}_{hole}$. The temperature dependence is assumed to arise from various excitation terms that are not explicitly captured in the model. Using the oxidation reaction, charge neutrality condition, and charge disproportionation reaction, a self-consistent approach was employed to predict δ and parameterize $\Delta H^{\circ}_{Ox}$ as a function of hole concentration ($h^{\cdot}$(T)), which represents the temperature-dependent concentration of $Fe^{4+}$ species:

$$\Delta H^{\circ}_{Ox} = \Delta H^{\circ}_{hole} - 0.03T + 14.76 \quad (8)$$

$$\Delta H^{\circ}_{hole} = 12659h^{\cdot}(T)^2 - 3187h^{\cdot}(T) \quad (9)$$

The solid-state vibrational entropy, $S^{vib}_{solid}(T)$, was calculated using three Einstein vibrational modes per oxygen atom, with an Einstein temperature of 500 K. [28] In addition, six Einstein modes were considered for two oxygen atoms, and the resulting entropy was divided by two to obtain the contribution per oxygen atom. [28] The gas-phase $O_2$ entropy, $S^{\circ}_{O2,gas}(T)$, was calculated using the Shomate equation with thermodynamic data obtained from the NIST Chemistry WebBook [39] and was divided by two to obtain the entropy per oxygen atom. The standard oxidation entropy was then calculated as

$$\Delta S^{\circ}_{Ox} = S^{vib}_{solid}(T) - \frac{1}{2}S^{\circ}_{O2,gas}(T) = \frac{1}{2}[6s_E(500K,\ T)] - \frac{1}{2}S^{\circ}_{O2,gas}(T) \quad (10)$$

$$s_E(\theta_E, T) = R\left[\frac{x}{e^x - 1} - \ln(1 - e^{-x})\right] = R\left[\frac{x/2}{\tanh(x/2)} - \ln(2\sinh(x/2))\right] \quad (11)$$

$$x = \frac{\theta_E}{T}, \theta_E = 500\ K \quad (12)$$

Equation 13 corresponds to the hydride formation reaction, a recently proposed alternative protonation pathway.[35] In this mechanism, a proton initially bonded to an oxygen lattice site ($OH^{\cdot}_O$) migrates to a neighboring oxygen vacancy ($V^{\cdot\cdot}_O$), resulting in the formation of a hydride species ($H^{\cdot}_O$).[35]

$$V^{\cdot\cdot}_O + 2Fe^{\times}_{Fe} + OH^{\cdot}_O \rightleftharpoons O^{\times}_O + 2Fe^{\cdot}_{Fe} + H^{\cdot}_O \quad (13)$$

$$K_{hydride} = \frac{(3-\delta-[OH^{\cdot}_O]-[H^{\cdot}_O])[Fe^{\cdot}_{Fe}]^2[H^{\cdot}_O]}{\delta[Fe^{\times}_{Fe}]^2[OH^{\cdot}_O]} = exp^{-(\Delta H^{\circ}_{hydride} - T\Delta S^{\circ}_{hydride})/RT} \quad (14)$$

This process is accompanied by the oxidation of two $Fe^{3+}$ ions ($Fe^{\times}_{Fe}$) to two $Fe^{4+}$ ions ($Fe^{\cdot}_{Fe}$). As in the oxidation reaction, hole transfer plays a critical role, and iron continues to serve as the primary hole-trapping center through the formation of $Fe^{4+}$ species ($Fe^{\cdot}_{Fe}$).

$$2Fe^{\times}_B \rightleftharpoons Fe'_B + Fe^{\cdot}_B \quad (15)$$

$$K_i = \frac{[Fe'_B][Fe^{\cdot}_B]}{[Fe^{\times}_B]^2} = exp^{-(\Delta H^{\circ}_i - T\Delta S^{\circ}_i)/RT} \quad (16)$$

In the present model, the nonconfigurational parts of the entropy changes associated with hydride formation ($\Delta S^{\circ}_{hydride}$) and charge disproportionation ($\Delta S^{\circ}_i$) were assumed to be zero. The generation and redistribution of these holes are further governed by the charge disproportionation reaction (Equation 15), in which two $Fe^{3+}$ ions ($Fe^{\times}_{Fe}$) transform into one $Fe^{2+}$ ($Fe'_{Fe}$) and one $Fe^{4+}$ ($Fe^{\cdot}_{Fe}$). This process provides a thermodynamically viable pathway for accommodating changes in the local redox environment and helps buffer the system's electronic charge during protonation events.[36]

Table 1. Total energies of $H_O^{\bullet}$ and $OH_O^{\bullet}$ at different sites, corresponding to hydride formation reactions involving $Fe^{3+}$ ($Fe_{Fe}^{\times}$) and $Fe^{4+}$ ($Fe_{Fe}^{\cdot}$) ions.

| $H_O^{\bullet}$ site | Total energy (eV) |
|---|---|
| Site 1 | -254.24 |
| Site 2 | -254.50 |
| $OH_O^{\bullet}$ site | Total energy (eV) |
| Site 1 | -256.57 |
| Site 2 | -256.08 |
| Site 3 | -257.17 |
| Site 4 | -255.65 |
| Site 5 | -256.26 |

Hydride formation enthalpies ($\Delta H_{hydride}^{DFT}$) were calculated using DFT and defined as the total energy difference between hydride ($H_O^{\cdot}$) and hydroxyl ($OH_O^{\cdot}$) species in various local atomic environments, as summarized in Table 1. The calculations were performed on a BCFZ supercell with a composition of $Ba_8Co_3Fe_3Zr_2O_{22}H$. The same total number of electrons was used for both the hydroxyl and hydride configurations because the charge change associated with the oxidation of two $Fe^{3+}$ ions to $Fe^{4+}$ was compensated by the change in the formal charge of hydrogen from +1 in hydroxyl to −1 in hydride, resulting in no net change in the total charge of the system. Due to the limited number of oxygen vacancies, only two $H_O^{\cdot}$ configurations were considered, while five randomly selected $OH_O^{\cdot}$ configurations were sampled. The calculated $\Delta H_{hydride}^{DFT}$ values range from 1.15 to 2.93 eV depending on the local configuration, corresponding to hydride formation reactions involving $Fe^{3+}$ ($Fe_{Fe}^{\times}$) and $Fe^{4+}$ ($Fe_{Fe}^{\cdot}$) ions. From Table 1, the lowest-energy hydride ($H_O^{\cdot}$) and hydroxyl ($OH_O^{\cdot}$) configurations were used to obtain a DFT hydride formation enthalpy, $\Delta H_{hydride}^{DFT}$, of 2.67 eV. Because this DFT reaction energy contains the hole-interaction contribution associated with the electronic state of the DFT supercell, this contribution was removed before incorporating the DFT energy into the thermodynamic model. The corrected reference hydride formation enthalpy was therefore defined as

$$\Delta H_{hydride}^{\circ} = \Delta H_{hydride}^{DFT} - \Delta H_{hole}^{\circ}(h_{DFT,2h}),$$

where $\Delta H_{hole}^{\circ}(h_{DFT,2h})$ is the hole-interaction contribution evaluated at the hole concentration corresponding to the two-hole state represented in the DFT hydride formation reaction. The hole-interaction contribution corresponding to the equilibrium hole concentration ($\Delta H_{hole}^{\circ}$) was then explicitly included. The resulting modified van't Hoff expression for hydride formation is:

$$K_{hydride} = exp^{-(\Delta H_{hydride}^{\circ} + \Delta H_{hole}^{\circ})/RT}$$

$$= exp^{-(\Delta H_{hydride}^{DFT} - \Delta H_{hole}^{\circ}(h_{DFT,2h}) + \Delta H_{hole}^{\circ})/RT} \quad (17)$$

To calculate $\Delta H_i^{\circ}$ in Equation (16), we first determined the hydride formation enthalpy ($\Delta H_{hydride}^{\circ}$) for the reaction between $Fe^{2+}$ ($Fe_{Fe}^{\prime}$) and $Fe^{3+}$ ($Fe_{Fe}^{\times}$) using DFT (Equation (18)).

$$V_O^{\cdot\cdot} + 2Fe_{Fe}^{\prime} + OH_O^{\cdot} \rightleftharpoons O_O^{\times} + 2Fe_{Fe}^{\times} + H_O^{\cdot} \quad (18)$$

The value of $\Delta H_i^{\circ}$ was then obtained using $\Delta H_{hydride}^{\circ}$ from Equations (13) and (18). The calculated $\Delta H_{hydride}^{\circ}$ for different $OH_O^{\cdot}$ and $H_O^{\cdot}$ configurations in BCFZ ranges from 0.30 to 1.71 eV (Table 2). Calculations were performed on a BCFZ supercell ($Ba_8Co_3Fe_3Zr_2O_{20}H$). As in the DFT calculations for Equation (13), the same total number of electrons was maintained for both the hydroxyl and hydride configurations. Based on the number of oxygen vacancies, four $H_O^{\cdot}$ configurations were considered, along with five randomly sampled $OH_O^{\cdot}$ configurations. These correspond to hydride formation reactions involving $Fe^{2+}$ ($Fe_{Fe}^{\prime}$) and $Fe^{3+}$ ($Fe_{Fe}^{\times}$) ions. The most stable total energy of $H_O^{\cdot}$ and $OH_O^{\cdot}$ were used for calculating $dH_{hydride}$ and gave a value of 1.44 eV.

Table 2. Total energies of $H_O^{\bullet}$ and $OH_O^{\bullet}$ at different sites, corresponding to hydride formation reactions involving $Fe^{2+}$ ($Fe_{Fe}^{\prime}$) and $Fe^{3+}$ ($Fe_{Fe}^{\times}$) ions

| $H_O^{\bullet}$ site | Total energy (eV) |
|---|---|
| Site 1 | -245.89 |
| Site 2 | -246.08 |
| Site 3 | -246.16 |
| Site 4 | -245.90 |
| $OH_O^{\bullet}$ site | Total energy (eV) |
| Site 1 | -247.41 |
| Site 2 | -247.57 |
| Site 3 | -247.60 |
| Site 4 | -246.89 |
| Site 5 | -246.46 |

Collectively, the four reactions including hydration, oxidation, hydride formation, and charge disproportionation govern the defect chemistry and the relative populations of hydroxyl ($OH_O^{\cdot}$) and hydride ($H_O^{\cdot}$) species in BCFZY systems, which in turn dictate the temperature and $pO_2$-dependent behavior of proton conductivity.

## 2.6. Thermodynamic Boundary Conditions

For the gas partial pressures used in the thermodynamic defect chemistry calculations, we adopted for almost all the results the typical PCFC cathode operating conditions of $pO_2 = 0.206$ atm and $pH_2O = 0.0296$ atm.[40–42] Additionally, in Sec. 3.1.3 we validated our results against experimental data obtained under low-$pO_2$ conditions. This choice is because the only available proton conductivity data that allow for a direct comparison of Y-doping effects across BCFZ, BCFZY0.1, and BCFY were obtained under reduced $pO_2$ conditions. To simulate these experimental conditions, thermodynamic boundary conditions were established based on hydrogen permeation experiments reported in the literature.[13] Since these experimental conductivity measurements were performed under steady-state rather than equilibrium conditions, it is necessary to approximate the thermodynamic state of the system. However, accurately estimating the effective hydrogen and oxygen partial pressures at the sample surface is challenging, as they cannot be measured directly.

To build a thermodynamic approximation to the steady state conditions we defined a closed system volume of 0.035 liters, corresponding to the feed side of the experimental tube.[13] The total gas pressure was fixed at 1 atm, and the total number of gas moles was assumed to remain constant throughout the simulation. The initial gas composition in the system was set to $H_2 : N_2 : O_2 = 5 : 95 : x$, reflecting the experimental conditions in which the sample was exposed to a 5 % $H_2$ / 95 % $N_2$ gas stream from a commercial cylinder. To account for potential oxygen impurities, leaks, and the release of $O_2$ from the BCFZY membrane, a variable $O_2$ mole fraction (x) was introduced. Initially, x was set to 0.0001 mol $O_2$ per mol of total gas to represent a feasible impurity level in the commercial gas cylinder. To further consider possible oxygen leakage in the chamber or $O_2$ release from the BCFZY membrane, a higher value of x = 0.1 was also examined.

Given the assumption that the initial gas was assumed dry in the system, consistent with the experimental conditions, we used these initial gas compositions to calculate the mole fractions of $H_2$, $O_2$, and $H_2O$ at equilibrium, applying the equilibrium constant (K) for the $H_2O$ formation reaction. From these mole fractions, we determined the effective partial pressures of $H_2$, $O_2$, and $H_2O$, which were then used as input values in the thermodynamic defect model.

## 3. Results and Discussion

### 3.1 Effect of Y doping

#### 3.1.1 Effect of Y doping on Proton diffusivities in BCFZY materials

Protons in perovskite oxides diffuse via two well-known mechanisms: the Grotthuss mechanism and the vehicle mechanism.[43] In the Grotthuss mechanism, proton transport occurs through a sequence of reorientations and hops between neighboring oxygen ions. The proton spends most of its time in a rotational state, coordinated to an oxygen ion, and occasionally hops to an adjacent oxygen ion to form a new covalent bond. In contrast, the vehicle mechanism involves the migration of a proton together with a carrier species, which is an $O^{2-}$ ion. Our AIMD simulations suggest that the protons in BCFZY follow the Grotthuss mechanism when present as OH and we assume they hop with a vacancy vehicle mechanism for Ho (see below for more discussion).

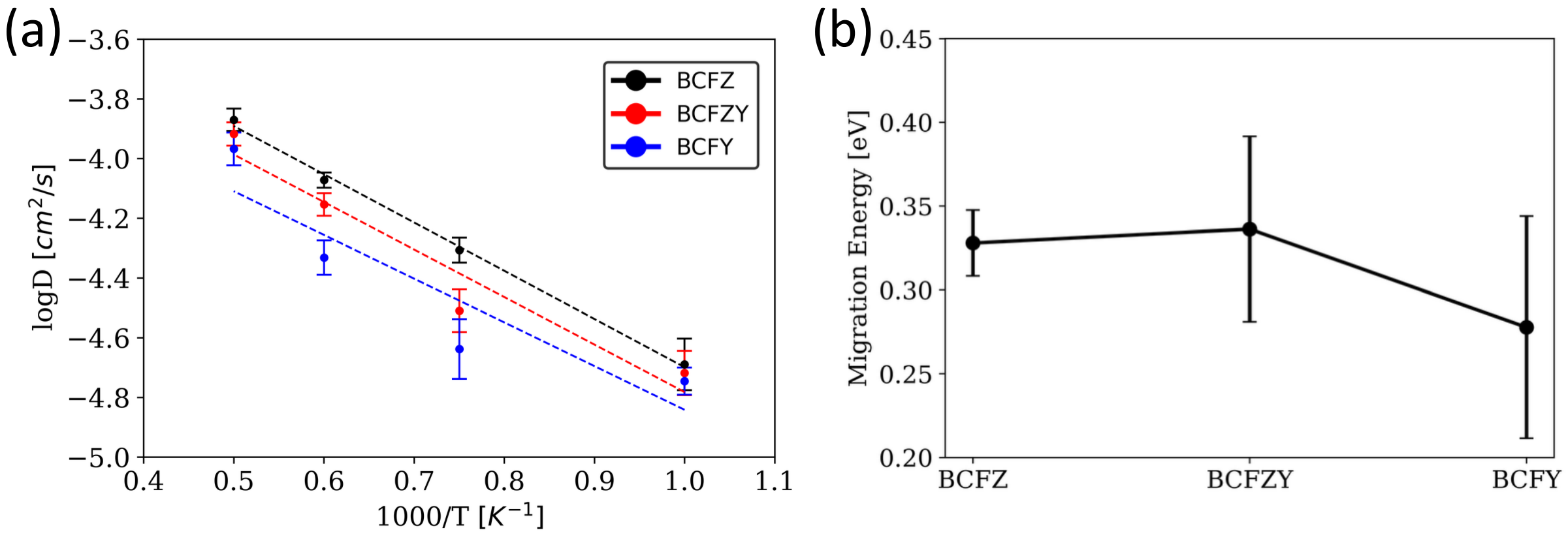

Figure 1. (a) Tracer diffusivity and $E_m$ (b) of proton in BCFZ, BCFZY0.1 and BCFY.

AIMD simulations were performed to quantify the influence of Y doping on proton diffusion in BCFZ, BCFZY0.1, and BCFY. The results, summarized in Fig. 1, suggest that Y doping has a relatively minor effect on the proton tracer diffusion coefficient ($D_H$) over the simulated temperature range of 1000 – 2000 K. The calculated migration energies are $0.33 \pm 0.02$ eV for BCFZ, $0.34 \pm 0.06$ eV for BCFZY0.1, and $0.28 \pm 0.07$ eV for BCFY, suggesting that Y substitution does not significantly affect the energy barrier for proton transport.

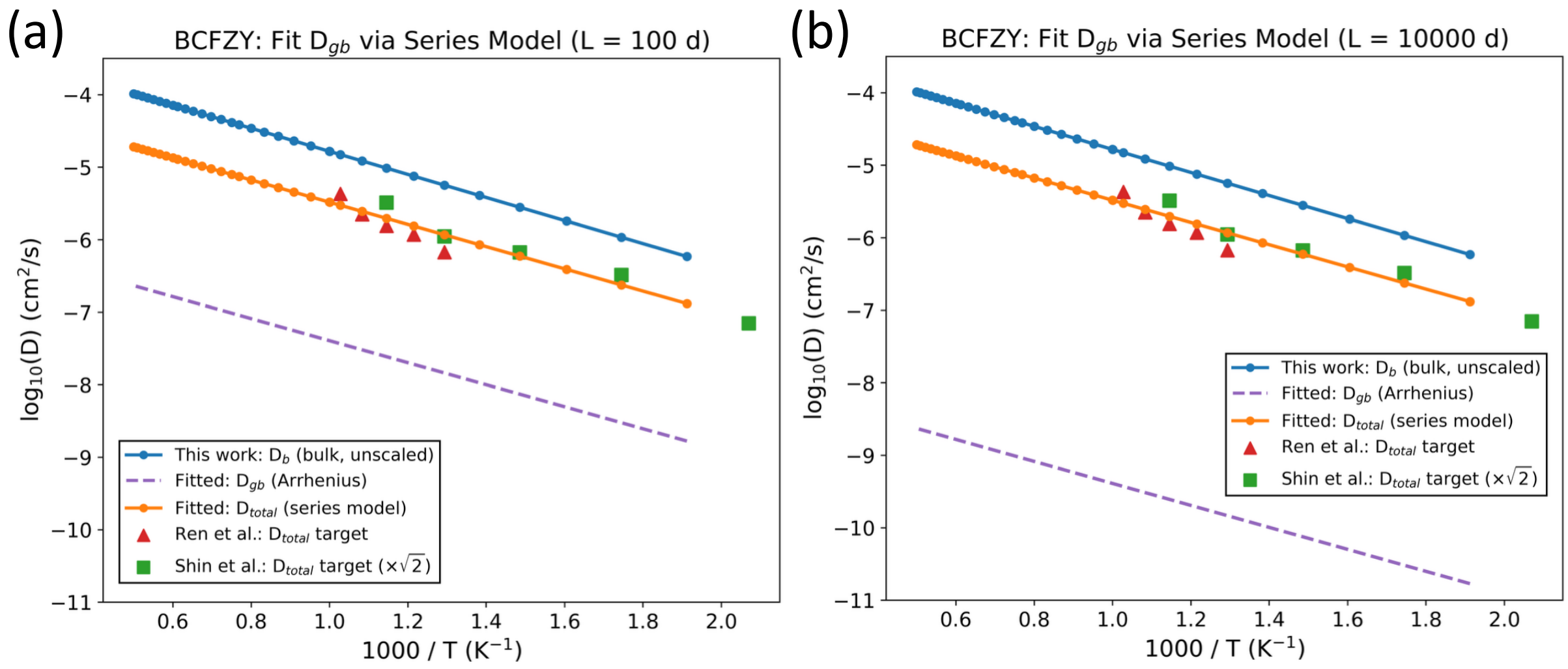

Figure 2. Series-model fitting of proton diffusivity in BCFZY0.1: (a) L=100d; (b) L = 10,000d

Grain boundary effects on proton diffusivity were incorporated using a series model fitted to experimental data (see Sec. 2.3). The results of the series-model fitting of proton diffusivity in BCFZY0.1 are shown in Figure 2. The bulk diffusivity was calculated using Equation 2. Both $L = 100d$ (Figure 2a) and $L = 10{,}000d$ (Figure 2b) cases are presented. Based on literature reports for grain boundaries in $BaZr_{0.9}Y_{0.1}O_{3-\delta}$, the grain boundary thickness (estimated using the brick-layer model) ranges from ~3.5 to 17.0 nm.[44] The reported grain sizes for some of the experimental samples used for comparison here were approximately 3 μm, 10 μm, and 18 μm for BCFZ, BCFZY0.1, and BCFY, respectively.[14] Considering these values, the ratio between grain size ($L$) and grain boundary thickness ($d$) was chosen in the range $L = 100 – 10{,}000d$. For both $L = 100d$ and $L = 10{,}000d$ cases, the migration energies ($E_m$) are 0.316 eV for bulk proton diffusivity, 0.301 eV for grain boundary diffusivity, and 0.304 eV for total proton diffusivity, indicating that the migration energy for proton diffusion is lower at the grain boundary than in the bulk. The pre-exponential factors for bulk and total diffusivity are $A_b = 6.43 \times 10^{-4}$ cm²/s and $A_{\text{total}} = 1.12 \times 10^{-4}$ cm²/s, respectively, and are independent of the assumed $L$ value. In contrast, the pre-exponential factor for grain boundary diffusivity, $A_{gb}$, depends strongly on $L$, decreasing from $1.33 \times 10^{-6}$ cm²/s for $L = 100d$ to $1.35 \times 10^{-8}$ cm²/s for $L = 10{,}000d$. Thus, the pre-exponential factor for grain boundary diffusivity is approximately 2–4 orders of magnitude lower than that for bulk diffusivity. This difference is consistent with literature reports suggesting grain boundary oxygen conductivity prefactors to be two to four orders of magnitude lower than bulk values. [37,38]

### 3.1.2 Effect of Y doping on proton concentration in BCFZY materials

The proton concentration in BCFZY materials was evaluated using a thermodynamic defect model approach, as described in Section 2.5. The model includes both conventional hydration mechanisms and a recently proposed hydride formation pathway. Proton incorporation was considered through two distinct reactions: the conventional hydration reaction that forms hydroxyl groups ($OH_O^{\cdot}$), and the hydride formation reaction, which creates protonic defects on oxygen lattice sites ($H_O^{\cdot}$).

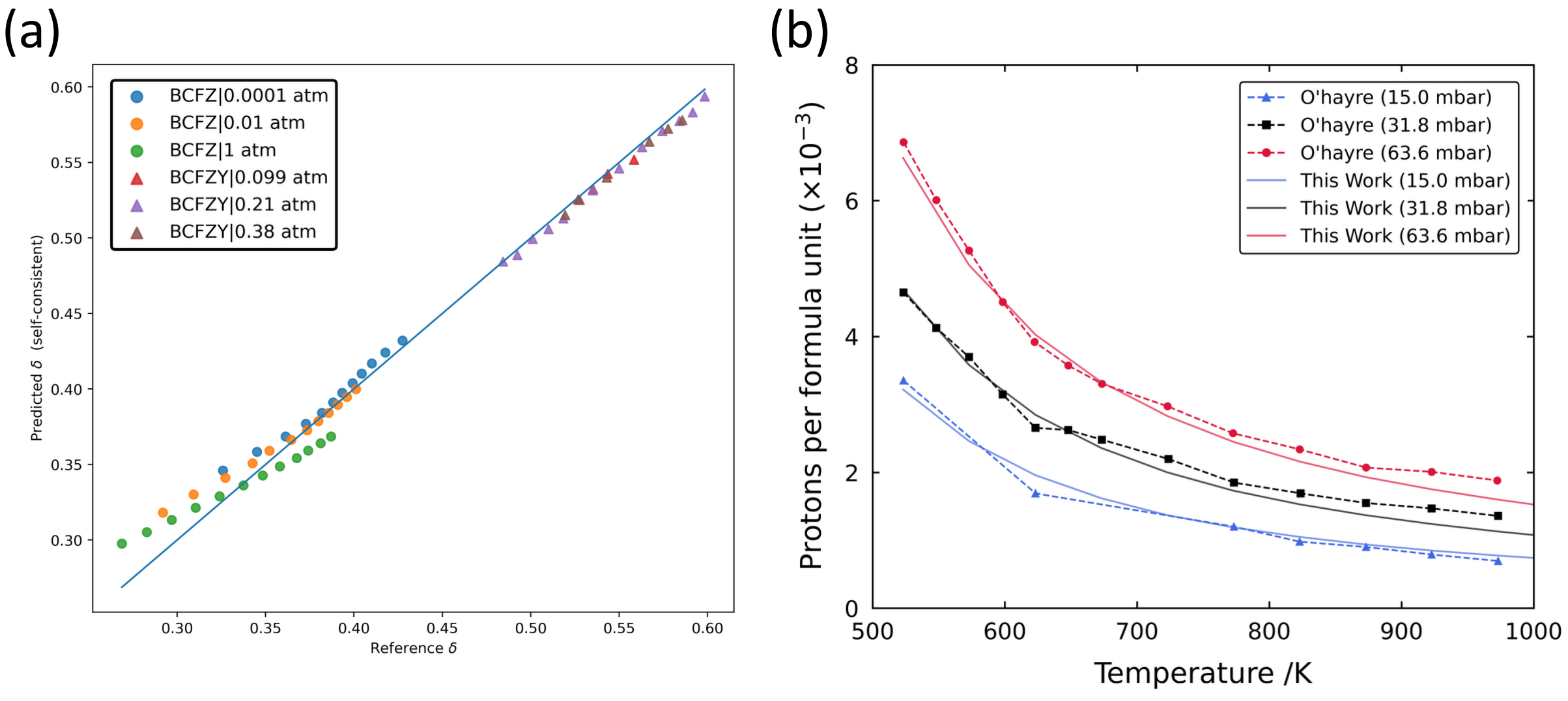

Figure 3. Results of fitted $\delta$ (a) and proton concentration (b) using thermodynamic defect model

To calculate the proton concentrations ($[OH_O^{\cdot}]$ and $[H_O^{\cdot}]$), it is necessary to determine $\Delta H_{Ox}^{\circ}$ and $\Delta S_{Ox}^{\circ}$, since these values are not available for BCFZY materials. In contrast, the hydration parameters ($\Delta H_{hydrat}^{\circ}$ and $\Delta S_{hydrat}^{\circ}$) can be obtained from the literature. The value of $\Delta S_{Ox}^{\circ}$ was calculated using the Einstein model for vibrational entropy (Section 2.5). To obtain $\Delta H_{Ox}^{\circ}$, we expressed it as a polynomial function of hole concentration and temperature, and determined the coefficients through an iterative fitting procedure. Specifically, the coefficients were optimized to ensure that the predicted oxygen nonstoichiometry (δ) closely matches reference δ values. [24,25] The full code used for the thermodynamic defect modeling and fitting procedure is publicly available in an online repository. [45] Figure 3 (a) presents a parity plot comparing the reference and predicted δ values for BCFZ and BCFZY under different oxygen partial pressures. The predicted δ shows good agreement with the reference data (MAE = 0.0072, RMSE = 0.0098, $R^2$ = 0.99). Figure 3 (b) shows the total proton concentration in BCFZY0.1, calculated from the defect model equations (Section 2.5) using the fitted $\Delta H_{Ox}^{\circ}$ expression. The calculated total proton concentration ($[OH_O^{\cdot}] + [H_O^{\cdot}]$) agrees well with the reference data measured under three different water partial pressures (15 mbar, 31.8 mbar, and 63.6 mbar) at a fixed oxygen partial pressure of 209 mbar. As this is the only available dataset reporting proton concentrations under PCFC operating conditions, it was used as the primary reference for validating our model.

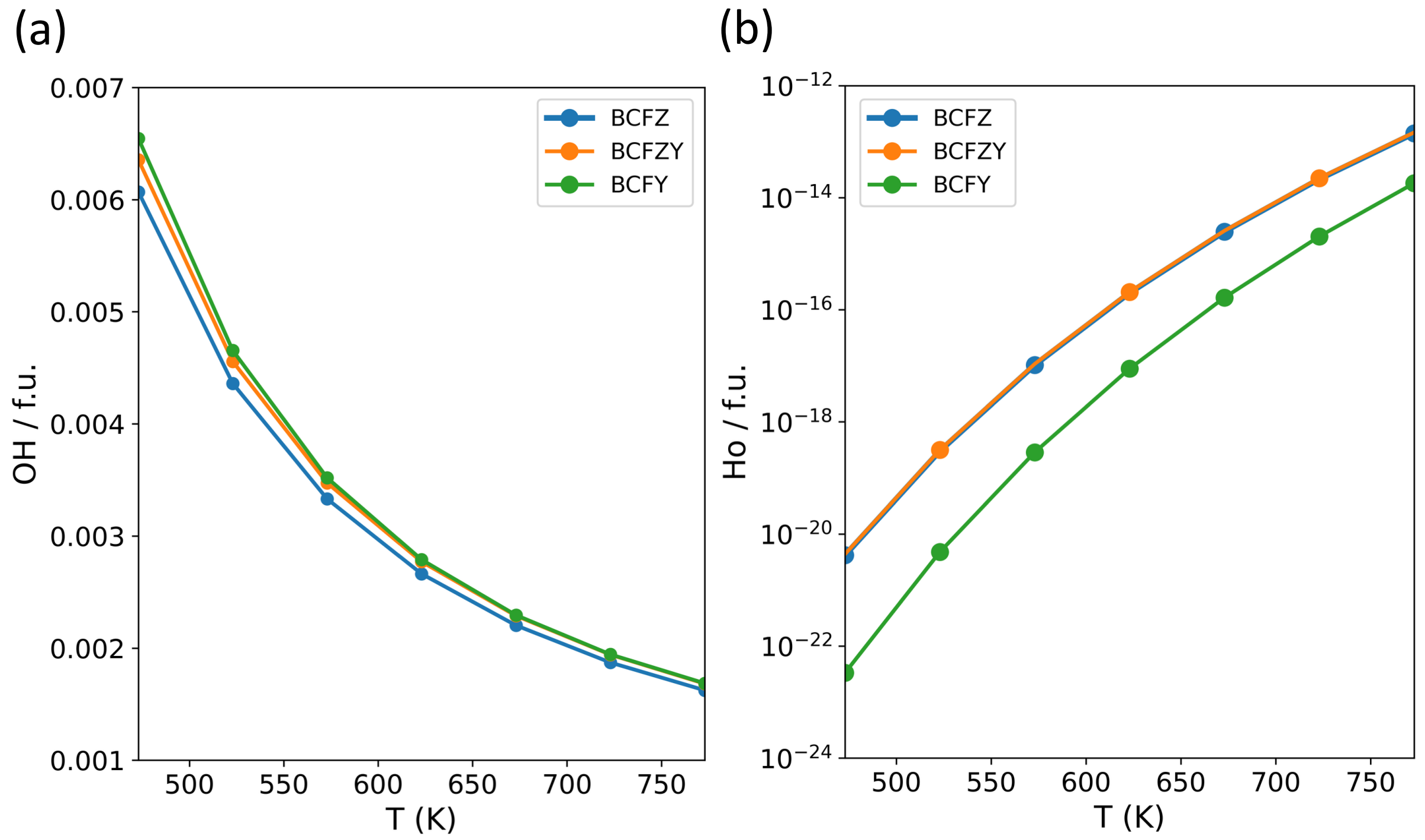

Figure 4. Calculated proton concentration of BCFZ, BCFZY0.1, BCFY (a) $OH_O^{\cdot}$ (b) $H_O^{\cdot}$ ($pO_2$ = 0.206 atm, $pH_2O$ = 0.0296 atm)

Figure 4 presents the temperature-dependent proton concentrations for BCFZ, BCFZY0.1, and BCFY under PCFC operating gas conditions ($pO_2$ = 0.206 atm, $pH_2O$ = 0.0296 atm), separated into contributions from $OH_O^{\cdot}$ and $H_O^{\cdot}$ species. As temperature increases, the concentration of hydroxyl protons ($OH_O^{\cdot}$) decreases due to the exothermic nature of hydration combined with the negative entropy change, which thermodynamically favors proton incorporation at lower temperatures.[25] In contrast, the concentration of hydride species ($H_O^{\cdot}$) increases with temperature. This trend is consistent with the positive formation enthalpy calculated by DFT for the hydride formation reaction, indicating that hydride concentrations increase at elevated temperatures despite remaining much lower than hydroxyl concentrations.

The incorporation of Y leads to increased oxygen vacancy concentrations, as discussed in Section 2.1, which in turn enhances the potential for proton uptake. However, the total proton concentration exhibits only modest variation with Y content across the investigated compositions. These trends suggest that while Y doping facilitates the structural conditions necessary for proton incorporation (i.e., by increasing δ), the actual proton concentrations are additionally controlled by the thermodynamic tendencies of the two protonation mechanisms.

Under the PCFC operating conditions examined in this study, the concentration of hydride species ($H_O$) is negligible and therefore has little effect on the total proton concentration. However, hydride species become increasingly important at elevated temperatures and under highly reducing conditions, particularly at low $pO_2$ or low oxygen chemical potential.

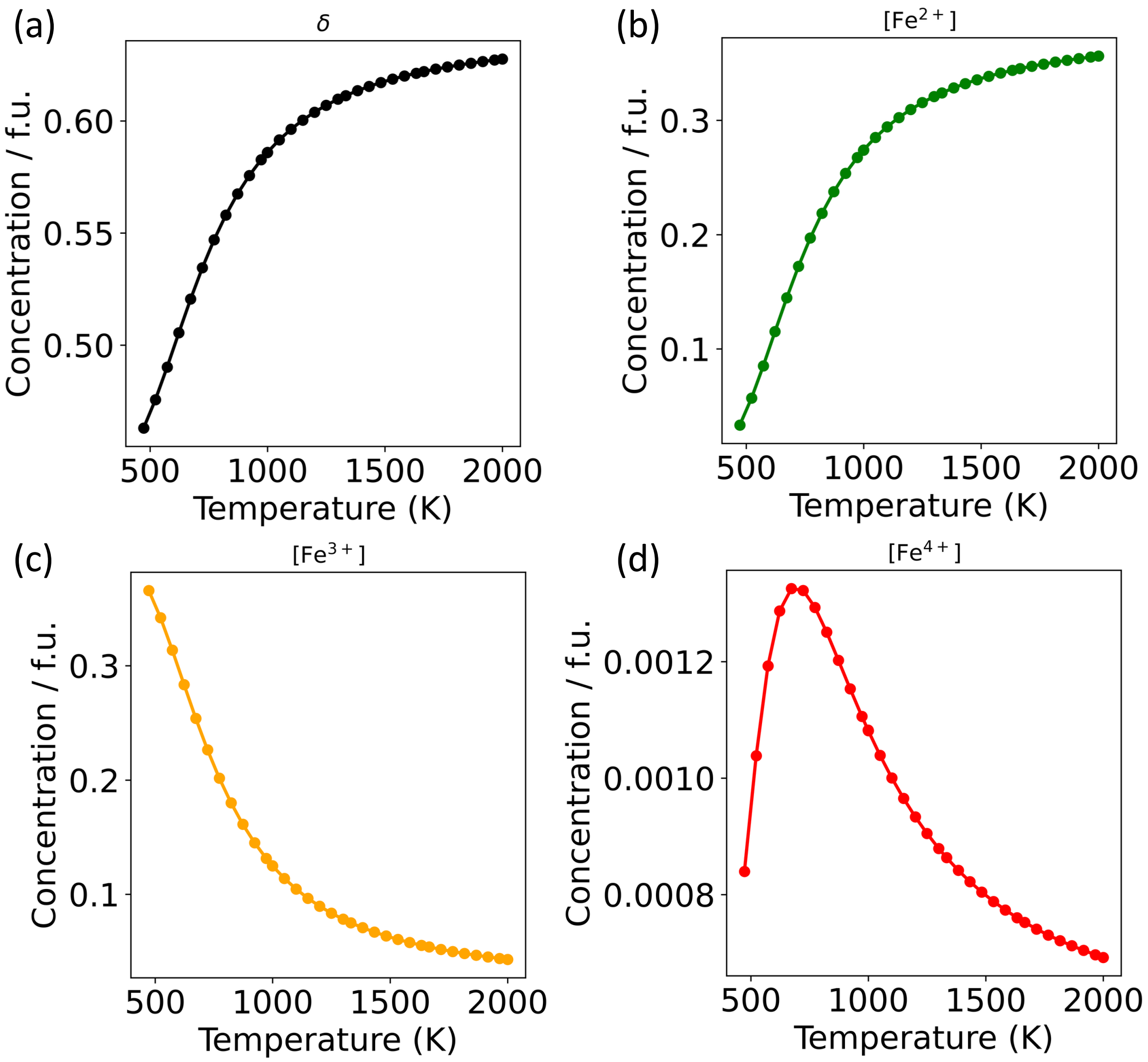

Figure 5. Calculated concentration of Fe ions and oxygen non-stoichiometry in BCFZY0.1 ($pO_2$ = 0.206 atm, $pH_2O$ = 0.0296 atm)

Figure 5 presents the calculated concentrations of Fe ions and oxygen non-stoichiometry in BCFZY0.1. As temperature increases, the oxygen non-stoichiometry (δ) increases due to oxygen vacancy formation, which is coupled with a shift in iron oxidation states: $Fe^{2+}$ concentration increases while $Fe^{3+}$ decreases, and the hole concentration ($Fe^{4+}$) increases up to approximately 673K and then decreases at higher temperatures. This nonmonotonic behavior can be understood from the charge neutrality condition, which couples the concentrations of hydroxyl defects, oxygen vacancies, and Fe oxidation states. At lower temperatures, the decrease in hydroxyl concentration contributes to an increase in $Fe^{4+}$ concentration, whereas at higher temperatures, the balance between oxygen vacancy formation and changes in Fe oxidation states leads to a decrease in $Fe^{4+}$ concentration. The contribution of hydride species to charge neutrality is negligible due to their extremely low concentrations under these conditions. This redox behavior is constrained by the charge neutrality condition (Section 2.5), which determines the feasible range of δ: approximately 0.2–0.6 for BCFZ, 0.25–0.65 for BCFZY0.1, and 0.3–0.7 for BCFY. Proton concentration can also influence the feasible range of δ through its contribution to charge neutrality. However, under the PCFC operating gas conditions considered in this study, this effect is negligible due to the relatively low proton concentration (Fig. 4).

### 3.1.3 Effect of Y doping on proton conductivities in BCFZY materials

Proton conductivities ($\sigma_H$) of BCFZ, BCFZY0.1, and BCFY under PCFC operating gas conditions ($pO_2$ = 0.206 atm, $pH_2O$ = 0.0296 atm) were calculated using the Nernst–Einstein relation:

$$\sigma_H = \frac{Z^2 e^2 c_H D_H}{kT} \quad (19)$$

where $c_H$ is the total proton concentration estimated from thermodynamic defect (Section 2.5) under the same PCFC operating gas conditions, and $D_H$ is the total proton diffusivity calculated as described in Section 2.3.

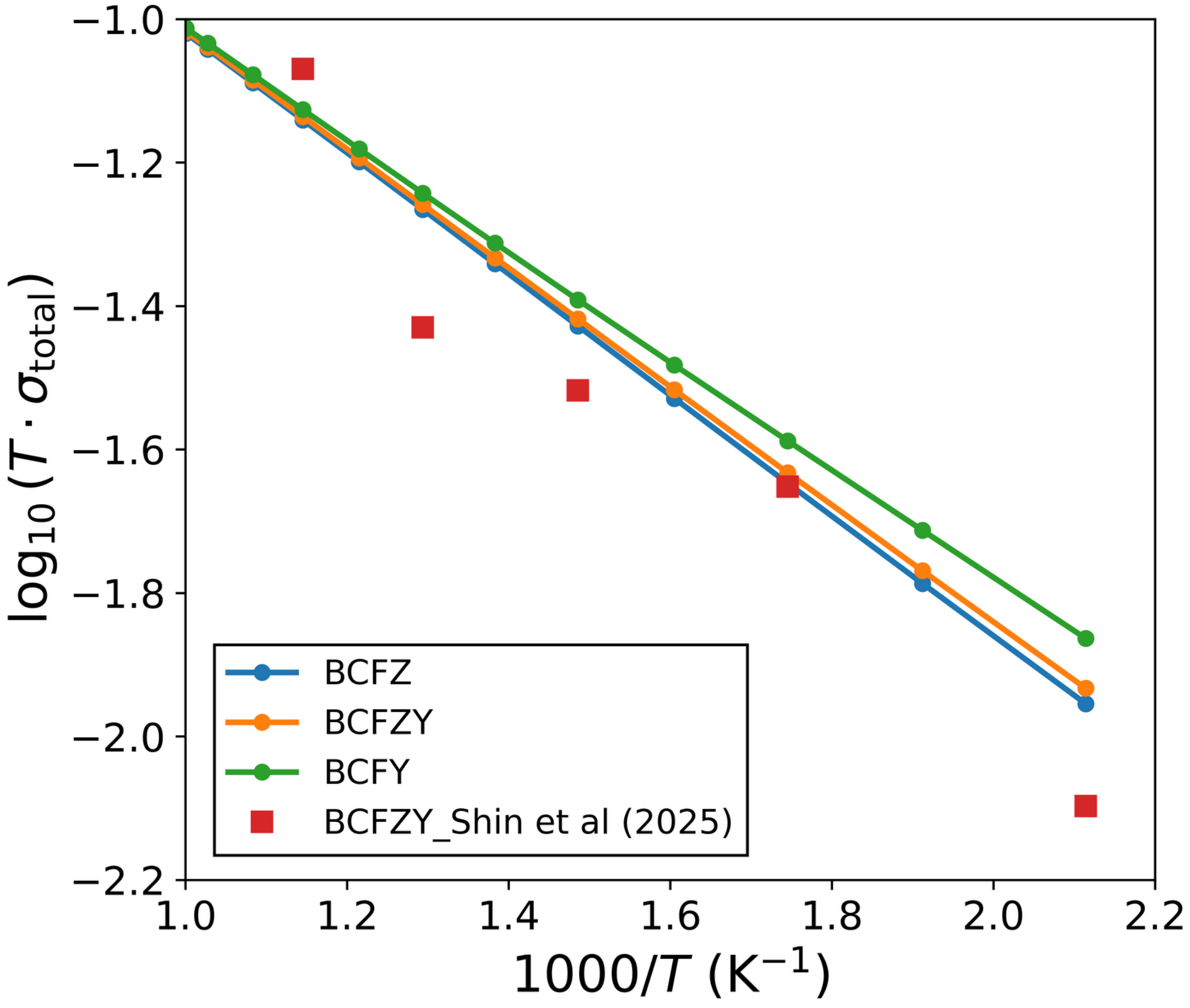

Figure 6. Proton conductivity ($\sigma_H$) of BCFZ, BCFZY0.1 and BCFY under PCFC operating gas conditions, compared with experimental reference.

From BCFZ to BCFZY0.1 to BCFY, the proton concentration increases, with a larger increase observed between BCFZY0.1 and BCFY than between BCFZ and BCFZY0.1. (Figure 6) The activation energy for proton conductivity associated with hydride species is significantly higher than that associated with hydroxyl groups. For example, in BCFZY, the activation energy is 0.181 ± 0.055 eV for $OH_O^{\cdot}$ conduction and 1.999 ± 0.015 eV for $H_O^{\cdot}$ conduction. Arrhenius fits for all BCFZY compositions are summarized in Table 3.

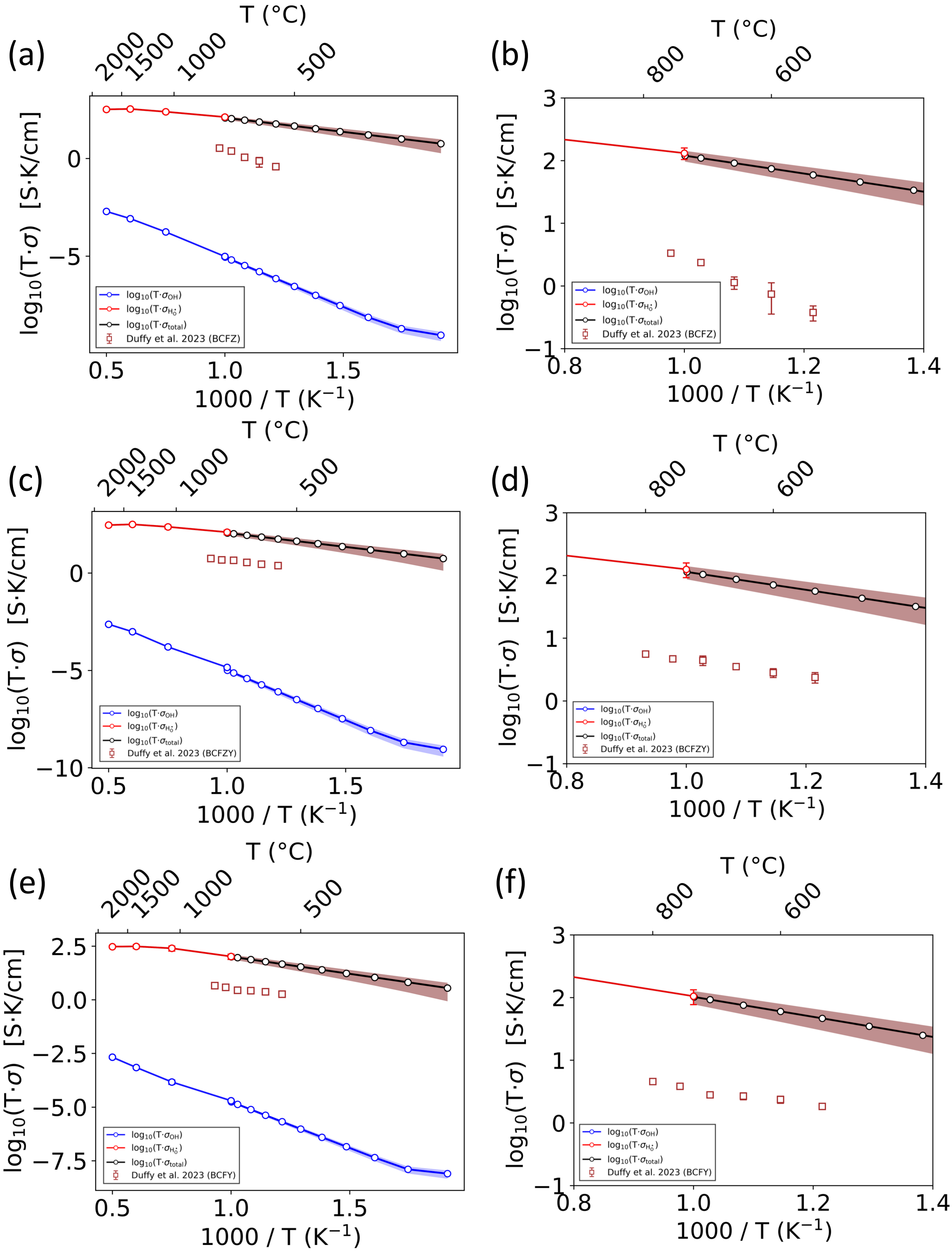

Figure 7. Calculated proton conductivity ($\sigma_H$) of BCFZ, BCFZY0.1, and BCFY under low $pO_2$ conditions, compared with experimental data. (a) BCFZ, (c) BCFZY0.1, and (e) BCFY over the full temperature range. Panels (b), (d), and (f) show enlarged views of the experimental temperature range for BCFZ, BCFZY0.1, and BCFY, respectively.

Additionally, we validated our results against experimental data obtained under low-$pO_2$ conditions (initial gas ratio: $H_2 : O_2 = 0.05 : 0.000001$). This comparison represents the only available basis for directly assessing Y-doping effects across BCFZ, BCFZY0.1, and BCFY. The calculated defect concentrations in BCFZY0.1 under these conditions are presented in Figure S10. Figure 7 presents the calculated proton conductivity under these conditions. Proton conductivity associated with hydride species ($H_O^{\cdot}$) dominates the total proton conductivity over the investigated temperature range. Furthermore, the activation energies obtained from experimental data are in good agreement with the modeling results for BCFZY0.1 and BCFY, whereas a larger discrepancy is observed for BCFZ. The difference in slope observed for BCFZ, compared to BCFZY0.1 and BCFY, may be attributed to microstructural effects that are mitigated by Y doping, or to differences in the experimental samples, as the BCFZ results deviate significantly from those of the Y-doped materials.

To account for possible oxygen leakage in the chamber or $O_2$ release from the BCFZY membranes, a slightly higher $pO_2$ condition ($H_2 : O_2 = 0.05 : 0.001$) was also considered (Figure S11). Under this condition, the proton conductivity associated with hydride species ($H_O^{\cdot}$), which dominates total proton conductivity, remains largely unchanged over the experimental temperature range compared with that under lower $pO_2$ case. In contrast, the proton conductivity associated with hydroxyl groups ($OH_O^{\cdot}$) increases, but its contribution to the total conductivity remains negligible.

Table 3. Arrhenius fitting parameters for protonic conductivity in all BCFZY compositions under two different conditions: (a) PCFC operating conditions and (b) reducing conditions.

(a) PCFC operating conditions

*($pO_2$ = 0.206 atm, $pH_2O$ = 0.0296 atm, 1000–2000 K)*

| Material | Species | $E_a$ (eV) | Pre-factor (S·K/cm) |
|---|---|---|---|
| BCFZ | $OH_O^{\cdot}$ | 0.162 ± 0.019 | (6.29 ± 0.90) × $10^{-1}$ |
| | $H_O^{\cdot}$ | 1.967 ± 0.018 | (4.67 ± 0.70) × 10 |
| BCFZY0.1 | $OH_O^{\cdot}$ | 0.181 ± 0.055 | (7.42 ± 3.03) × $10^{-1}$ |
| | $H_O^{\cdot}$ | 1.999 ± 0.015 | (4.68 ± 0.53) × 10 |
| BCFY | $OH_O^{\cdot}$ | 0.172 ± 0.066 | (7.33 ± 4.34) × $10^{-1}$ |
| | $H_O^{\cdot}$ | 2.439 ± 0.013 | (1.62 ± 0.16) × $10^{3}$ |

(b) Reducing conditions

*(initial gas ratio: $H_2$ : $O_2$ = 0.05 : 0.000001, 1000–2000 K)*

| Material | Species | $E_a$ (eV) | Pre-factor (S·K/cm) |
|---|---|---|---|
| **BCFZ** | $OH_O^{\cdot}$ | 0.894 ± 0.051 | (3.97 ± 1.49) × $10^{-1}$ |
| | $H_O^{\cdot}$ | 0.160 ± 0.035 | (9.18 ± 2.67) × $10^{2}$ |
| **BCFZY0.1** | $OH_O^{\cdot}$ | 0.890 ± 0.035 | (4.31 ± 1.12) × $10^{-1}$ |
| | $H_O^{\cdot}$ | 0.136 ± 0.045 | (7.01 ± 2.37) × $10^{2}$ |
| **BCFY** | $OH_O^{\cdot}$ | 0.794 ± 0.028 | (1.93 ± 0.48) × $10^{-1}$ |
| | $H_O^{\cdot}$ | 0.166 ± 0.049 | (8.51 ± 3.10) × $10^{2}$ |

## 4. Conclusion

In this study, we investigated the effect of Yttrium (Y) doping on proton conductivity in Ba (Fe, Co, Zr, Y) $O_{3-\delta}$ (BCFZY) materials, which are promising cathode candidates for protonic ceramic fuel cells (PCFCs). By decoupling the contributions of proton diffusivity and proton concentration, we aimed to clarify the mechanistic role of Y doping in tuning proton transport. Using ab initio molecular dynamics (AIMD) simulations, we observed that increasing Y content from BCFZ to BCFZY0.1 and BCFY has only a minor effect on proton tracer diffusivity. The calculated migration energies remain relatively unchanged across compositions, with values of 0.33 eV for BCFZ, 0.34 eV for BCFZY0.1, and 0.28 eV for BCFY, suggesting that Y doping does not significantly alter the diffusion barrier for proton transport. Thermodynamic defect modeling incorporating hydration and hydride formation reactions was employed to estimate temperature-dependent proton concentrations. These results revealed that Y doping modestly enhances proton incorporation, primarily due to the increase in oxygen vacancies introduced by the redox-inactive $Y^{3+}$ dopants substituting for $Zr^{4+}$. Calculated proton conductivities using the Nernst-Einstein relation under PCFC operating gas conditions reveal that the majority of proton transport arises from hydroxyl-type defects ($OH_O^{\cdot}$) rather than hydride species ($H_O^{\cdot}$). Additionally, our results were validated against experimental data obtained under low-$pO_2$ conditions (gas ratio: $H_2$ : $O_2$ = 0.05 : 0.000001). Under these conditions, proton conductivity is predominantly governed by hydride species ($H_O^{\cdot}$) over the investigated temperature range. When considered together with our previous results on oxygen transport [18], these findings reveal that Y doping has distinct effects on proton and oxygen conduction in BCFZY. Under PCFC operating gas conditions, Y substitution increases the oxygen vacancy concentration and modestly enhances proton incorporation, while proton diffusivity remains relatively insensitive to Y content, resulting in only a small change in proton conductivity. In contrast, our previous study suggested that although the increased oxygen vacancy concentration with Y doping favors oxygen transport, this effect is outweighed by the reduction in oxygen vacancy diffusivity, resulting in an overall decrease in oxygen conductivity. [18] Thus, we predict that increasing Y content does not uniformly enhance ionic transport but instead affects proton and oxygen conduction differently through its competing effects on defect concentration and mobility. Together, these findings provide a mechanistic understanding of how Y doping influences defect chemistry and both proton and oxygen transport in triple-conducting BCFZY cathode materials for PCFCs.

## Acknowledgements

This work was supported by the U.S. Department of Energy's National Energy Technology Laboratory (NETL), in part through a site support contract. Neither the United States Government nor any agency thereof, nor any of their employees, nor the support contractor or its employees, makes any warranty, express or implied, or assumes any legal liability or responsibility for the accuracy, completeness, or usefulness of any information, apparatus, product, or process disclosed, or represents that its use would not infringe privately owned rights. Reference to any specific commercial product, process, or service by trade name, trademark, manufacturer, or otherwise does not constitute or imply its endorsement, recommendation, or favoring by the United States Government or any agency thereof. The views and opinions expressed herein are those of the authors and do not necessarily reflect those of the United States Government or any agency thereof. Model refinement and writing was supported by the US Department of Energy (DOE), Office of Science, Basic Energy Sciences (BES), under award no. DE-SC0026102. Computations were performed on the Stampede3 system at the Texas Advanced Computing Center (TACC), The University of Texas at Austin, under allocation TG-MAT240071 from the Advanced Cyberinfrastructure Coordination Ecosystem: Services & Support (ACCESS) program, supported by National Science Foundation grants 2138259, 2138286, 2138307, 2137603, and 2138296. [46] Additional computational resources were provided by the Center for High Throughput Computing at the University of Wisconsin–Madison. [47] C.K., R.J., and D.M. thank Dr. Yueh-Lin Lee and Dr. Jun Meng for their advice on DFT calculations and for valuable discussions throughout this work.

## Conflict of Interest

The authors declare no conflicts or interests.

## Data Availability Statement

The data and codes that support the findings of this study are openly available in Figshare at https://doi.org/10.6084/m9.figshare.35015950, reference number [45].

## 5. References

[1] M. Singh, D. Zappa, E. Comini, Solid oxide fuel cell: Decade of progress, future perspectives and challenges, Int. J. Hydrogen Energy 46 (2021) 27643–27674. https://doi.org/10.1016/j.ijhydene.2021.06.020.

[2] Z. Zakaria, Z. Awang Mat, S.H. Abu Hassan, Y. Boon Kar, A review of solid oxide fuel cell component fabrication methods toward lowering temperature, Int. J. Energy Res. 44 (2020) 594–611. https://doi.org/10.1002/er.4907.

[3] D. Ding, X. Li, S.Y. Lai, K. Gerdes, M. Liu, Enhancing SOFC cathode performance by surface modification through infiltration, Energy Environ. Sci. 7 (2014) 552–575. https://doi.org/10.1039/c3ee42926a.

[4] W. Bian, W. Wu, B. Wang, W. Tang, M. Zhou, C. Jin, H. Ding, W. Fan, Y. Dong, J. Li, D. Ding, Revitalizing interface in protonic ceramic cells by acid etch, Nature 604 (2022) 479–485. https://doi.org/10.1038/s41586-022-04457-y.

[5] C. Duan, J. Tong, M. Shang, S. Nikodemski, M. Sanders, S. Ricote, A. Almansoori, R. O'Hayre, Readily processed protonic ceramic fuel cells with high performance at low temperatures, Science (1979). 349 (2015) 1321–1326. https://doi.org/10.1126/science.aab3987.

[6] S.M. Choi, H. An, K.J. Yoon, B.K. Kim, H.W. Lee, J.W. Son, H. Kim, D. Shin, H. Il Ji, J.H. Lee, Electrochemical analysis of high-performance protonic ceramic fuel cells based on a columnar-structured thin electrolyte, Appl. Energy 233–234 (2019) 29–36. https://doi.org/10.1016/j.apenergy.2018.10.043.

[7] E. Fabbri, D. Pergolesi, E. Traversa, Electrode materials: A challenge for the exploitation of protonic solid oxide fuel cells, Sci. Technol. Adv. Mater. 11 (2010). https://doi.org/10.1088/1468-6996/11/4/044301.

[8] C. Xia, Y. Mi, B. Wang, B. Lin, G. Chen, B. Zhu, Shaping triple-conducting semiconductor BaCo 0.4 Fe 0.4 Zr 0.1 Y 0.1 O 3-δ into an electrolyte for low-temperature solid oxide fuel cells, Nat. Commun. 10 (2019). https://doi.org/10.1038/s41467-019-09532-z.

[9] R. Ren, J. Sun, G. Wang, C. Xu, J. Qiao, W. Sun, Z. Wang, K. Sun, Rational design of Sr2Fe1.5Mo0.4Y0.1O6-δ oxygen electrode with triple conduction for hydrogen production in protonic ceramic electrolysis cell, Sep. Purif. Technol. 299 (2022). https://doi.org/10.1016/j.seppur.2022.121780.

[10] J.H. Duffy, N. Birkner, C. Kim, R. Jacobs, D. Morgan, S. Sharma, S.T. Misture, E.M. Kelder, H.W. Abernathy, K.S. Brinkman, Structural and thermodynamic analysis of triple conducting ceramic materials BaCo0.4Fe0.4Zr0.2−XYXO3−δ, J. Mater. Chem. A Mater. (2025). https://doi.org/10.1039/d4ta06417h.

[11] E. Gilardi, E. Fabbri, L. Bi, J.L.M. Rupp, T. Lippert, D. Pergolesi, E. Traversa, Effect of Dopant-Host Ionic Radii Mismatch on Acceptor-Doped Barium Zirconate Microstructure and Proton Conductivity, Journal of Physical Chemistry C 121 (2017) 9739–9747. https://doi.org/10.1021/acs.jpcc.7b02163.

[12] R. Merkle, M.F. Hoedl, G. Raimondi, R. Zohourian, J. Maier, Oxides with Mixed Protonic and Electronic Conductivity, Annu. Rev. Mater. Res. (2021). https://doi.org/10.1146/annurev-matsci-091819.

[13] J.H. Duffy, H.W. Abernathy, K.S. Brinkman, Tuning proton kinetics in BaCo0.4Fe0.4Zr0.2-XYXO3-δ triple ionic-electronic conductors via aliovalent substitution, J. Mater. Chem. A Mater. 11 (2023) 8929–8938. https://doi.org/10.1039/d3ta00654a.

[14] J.H. Duffy, Y. Meng, H.W. Abernathy, K.S. Brinkman, Surface and bulk oxygen kinetics of baco0.4fe0.4zr0.2−xyxo3−δ triple conducting electrode materials, Membranes (Basel). 11 (2021). https://doi.org/10.3390/membranes11100766.

[15] T. Hong, W. Lu, K. Ren, T. Liu, The two-fold diffusion process for proton uptake reaction in BCFZY e-/H + /O 2-triple conductor measured by electrical conductivity relaxation, (n.d.). https://doi.org/10.1007/s11581-020-03757-5/Published.

[16] P. Wang, D. Xu, J. Cheng, T. Hong, Proton uptake kinetics and electromotive force in BaCo 0.4 Fe 0.4 Zr 0.1 Y 0.1 O 3-δ cathode material with e − /O 2− /H + three mobile carriers for protonic ceramic fuel cells, (n.d.). https://doi.org/10.1007/s11581-021-03914-4/Published.

[17] H. Niu, Y. Jing, Y. Sun, L. Guo, N.R. Aluru, W. Li, J. Yang, X. Li, On the anomalous diffusion of proton in Y-doped BaZrO3 perovskite oxide, Solid State Ion. 376 (2022). https://doi.org/10.1016/j.ssi.2022.115859.

[18] C. Kim, R. Jacobs, J.H. Duffy, K.S. Brinkman, H.W. Abernathy, D. Morgan, Effects of Yttrium doping on Oxygen Conductivity in Ba (Fe, Co, Zr, Y) O3-δ Cathode Materials for Proton Ceramic Fuel Cells, Physical Chemistry Chemical Physics 28 (2026) 20805–20815. https://doi.org/10.1039/D5CP04074D.

[19] G. Kresse, J. Furthmü, Efficient iterative schemes for ab initio total-energy calculations using a plane-wave basis set, 1996.

[20] J.P. Perdew, K. Burke, M. Ernzerhof, Generalized Gradient Approximation Made Simple, 1996.

[21] V.I. Anisimov, F. Aryasetiawan, A.I. Lichtenstein, First-principles calculations of the electronic structure and spectra of strongly correlated systems: the LDA+ U method First-principles calculations of the electronic structure and spectra of strongly correlated systems: the LDA + U method, 1997.

[22] G. Kresse, D. Joubert, From ultrasoft pseudopotentials to the projector augmented-wave method, n.d.

[23] H.J. Monkhorst, J.D. Pack, Special points for Brillonin-zone integrations*, 1976.

[24] Y. Shin, M.D. Sanders, E. Truong, S. Majumder, B. Cladek, M. Walker, B. Ogbolu, R. Zhang, G.A. Evmenenko, Y.Y. Hu, M.J. Bedzyk, K. Page, S.M. Haile, R. O'Hayre, Thermodynamics, local structure, and transport of protons in triple-conducing oxide, BaCo0.4Fe0.4Zr0.1Y0.1O3-δ (BCFZY4411), Solid State Ion. 429 (2025). https://doi.org/10.1016/j.ssi.2025.116962.

[25] R. Zohourian, R. Merkle, J. Maier, Proton uptake into the protonic cathode material BaCo0.4Fe0.4Zr0.2O3-δ and comparison to protonic electrolyte materials, Solid State Ion. 299 (2017) 64–69. https://doi.org/10.1016/j.ssi.2016.09.012.

[26] W.G. Hoover, Canonical dynamics: Equilibrium phase-space distributions, 1985.

[27] S. Nose, Constant Temperature Molecular Dynamics Methods, Progress of Theoretical Physics Supplement (1991). https://doi.org/10.1143/PTPS.103.1/1894286.

[28] Y.L. Lee, J. Kleis, J. Rossmeisl, D. Morgan, Ab initio energetics of LaBO3 (001) (B=Mn, Fe, Co, and Ni) for solid oxide fuel cell cathodes, Phys. Rev. B Condens. Matter Mater. Phys. 80 (2009). https://doi.org/10.1103/PhysRevB.80.224101.

[29] J. Lee, N. Ohba, R. Asahi, First-principles prediction of high oxygen-ion conductivity in trilanthanide gallates Ln3GaO6, Sci. Technol. Adv. Mater. 20 (2019) 144–159. https://doi.org/10.1080/14686996.2019.1578183.

[30] C. Chen, D. Chen, Y. Gao, Z. Shao, F. Ciucci, Computational and experimental analysis of Ba0.95La 0.05FeO3-δ as a cathode material for solid oxide fuel cells, J. Mater. Chem. A Mater. 2 (2014) 14154–14163. https://doi.org/10.1039/c4ta01593b.

[31] John. Crank, The mathematics of diffusion, Clarendon Press, 1975.

[32] R. Ren, X. Yu, Z. Wang, C. Xu, T. Song, W. Sun, J. Qiao, K. Sun, Fluorination inductive effect enables rapid bulk proton diffusion in BaCo0.4Fe0.4Zr0.1Y0.1O3-δ perovskite oxide for high-activity protonic ceramic fuel cell cathode, Appl. Catal. B 317 (2022). https://doi.org/10.1016/j.apcatb.2022.121759.

[33] M.K. Hossain, T. Yamamoto, K. Hashizume, Isotopic effect of proton conductivity in barium zirconates for various hydrogen-containing atmospheres, J. Alloys Compd. 903 (2022). https://doi.org/10.1016/j.jallcom.2022.163957.

[34] C. Chen, S.P. Ong, A Universal Graph Deep Learning Interatomic Potential for the Periodic Table, 2022.

[35] H. Qi, Y.L. Lee, T. Yang, W. Li, W. Li, L. Ma, S. Hu, Y. Duan, G.A. Hackett, X. Liu, Positive Effects of H2O on the Hydrogen Oxidation Reaction on Sr2Fe1.5Mo0.5O6-δ-Based Perovskite Anodes for Solid Oxide Fuel Cells, ACS Catal. 10 (2020) 5567–5578. https://doi.org/10.1021/acscatal.9b05458.

[36] Y.-L. Lee, Y. Duan, D.C. Sorescu, W.A. Saidi, D. Morgan, T. Kalapos, W.K. Epting, G.A. Hackett, H.W. Abernathy, Defect Thermodynamic Modeling of Triple Conducting Perovskites (La,Ba)Fe 1-x M x O 3-δ for Proton-Conducting Solid-Oxide Cells , ECS Trans. 111 (2023) 1823–1838. https://doi.org/10.1149/11106.1823ecst.

[37] R. GERHARDT, A.S. NOWICK, Grain-Boundary Effect in Ceria Doped with Trivalent Cations: I, Electrical Measurements, Journal of the American Ceramic Society 69 (1986) 641–646. https://doi.org/10.1111/j.1151-2916.1986.tb07464.x.

[38] T. Chen, G.F. Harrington, J. Matsuda, K. Sasaki, D. Pham, E.L. Corral, N.H. Perry, Modifying Grain Boundary Ionic/Electronic Transport in Nano-Sr- and Mg- Doped LaGaO 3-δ by Sintering Variations , J. Electrochem. Soc. 166 (2019) F569–F580. https://doi.org/10.1149/2.0151910jes.

[39] P.J./ M.W.G. Linstrom, NIST Chemistry WebBook, NIST Standard Reference Database Number 69 (n.d.). https://doi.org/10.18434/T4D303.

[40] F. He, M. Liang, W. Wang, R. Ran, G. Yang, W. Zhou, Z. Shao, High-Performance Proton-Conducting Fuel Cell with B-Site-Deficient Perovskites for All Cell Components, Energy and Fuels 34 (2020) 11464–11471. https://doi.org/10.1021/acs.energyfuels.0c02370.

[41] S. Choi, T.C. Davenport, S.M. Haile, Protonic ceramic electrochemical cells for hydrogen production and electricity generation: Exceptional reversibility, stability, and demonstrated faradaic efficiency, Energy Environ. Sci. 12 (2019) 206–215. https://doi.org/10.1039/c8ee02865f.

[42] M. Shang, J. Tong, R. O'Hayre, A promising cathode for intermediate temperature protonic ceramic fuel cells: BaCo0.4Fe0.4Zr0.2O3-δ, RSC Adv. 3 (2013) 15769–15775. https://doi.org/10.1039/c3ra41828f.

[43] N. Wang, C. Tang, L. Du, R. Zhu, L. Xing, Z. Song, B. Yuan, L. Zhao, Y. Aoki, S. Ye, Advanced Cathode Materials for Protonic Ceramic Fuel Cells: Recent Progress and Future Perspectives, Adv. Energy Mater. 12 (2022). https://doi.org/10.1002/aenm.202201882.

[44] F. Iguchi, T. Tsurui, N. Sata, Y. Nagao, H. Yugami, The relationship between chemical composition distributions and specific grain boundary conductivity in Y-doped BaZrO3 proton conductors, Solid State Ion. 180 (2009) 563–568. https://doi.org/10.1016/j.ssi.2008.12.006.

[45] Data for "Effects of Yttrium doping on Proton Conductivity in Ba (Fe, Co, Zr, Y) O3-δ Cathode Materials for Proton Ceramic Fuel Cells" https://doi.org/10.6084/m9.figshare.35015950, (n.d.).

[46] T.J. Boerner, S. Deems, T.R. Furlani, S.L. Knuth, J. Towns, ACCESS: Advancing Innovation: NSF's Advanced Cyberinfrastructure Coordination Ecosystem: Services & Support, in: PEARC 2023 - Computing for the Common Good: Practice and Experience in Advanced Research Computing, Association for Computing Machinery, Inc, 2023: pp. 173–176. https://doi.org/10.1145/3569951.3597559.

[47] Center for High Throughput Computing, https://chtc.cs.wisc. edu/ (accessed April 09, 2025). DOI: 10.21231/GNT1-HW21., (n.d.).

Supplementary information for

# Effects of Yttrium doping on Proton Conductivity in Ba (Fe, Co, Zr, Y) $O_{3-\delta}$ Cathode Materials for Proton Ceramic Fuel Cells

Chiyoung Kim[a], Ryan Jacobs[a], Jun Meng[a], Jack H. Duffy[b], Kyle S. Brinkman[b], Yueh-Lin Lee[c], Harry W. Abernathy[c], Dane Morgan[a]

a. Department of Materials Science and Engineering, University of Wisconsin – Madison, Madison, Wisconsin 54706, United States
b. Department of Materials Science and Engineering, Clemson University, Clemson, South Carolina 29634, United States
c. National Energy Technology Laboratory, United States Department of Energy, Morgantown, West Virginia 26507, United States

Note that in the following references to figures and sections that are in the SI will have an "S" in front of the number, e.g., Figure S1 and Sec. S1. References to figures and sections without the "S" are to the main text.

## S1. Proton tracer diffusivity at different volume expansions in BCFZ

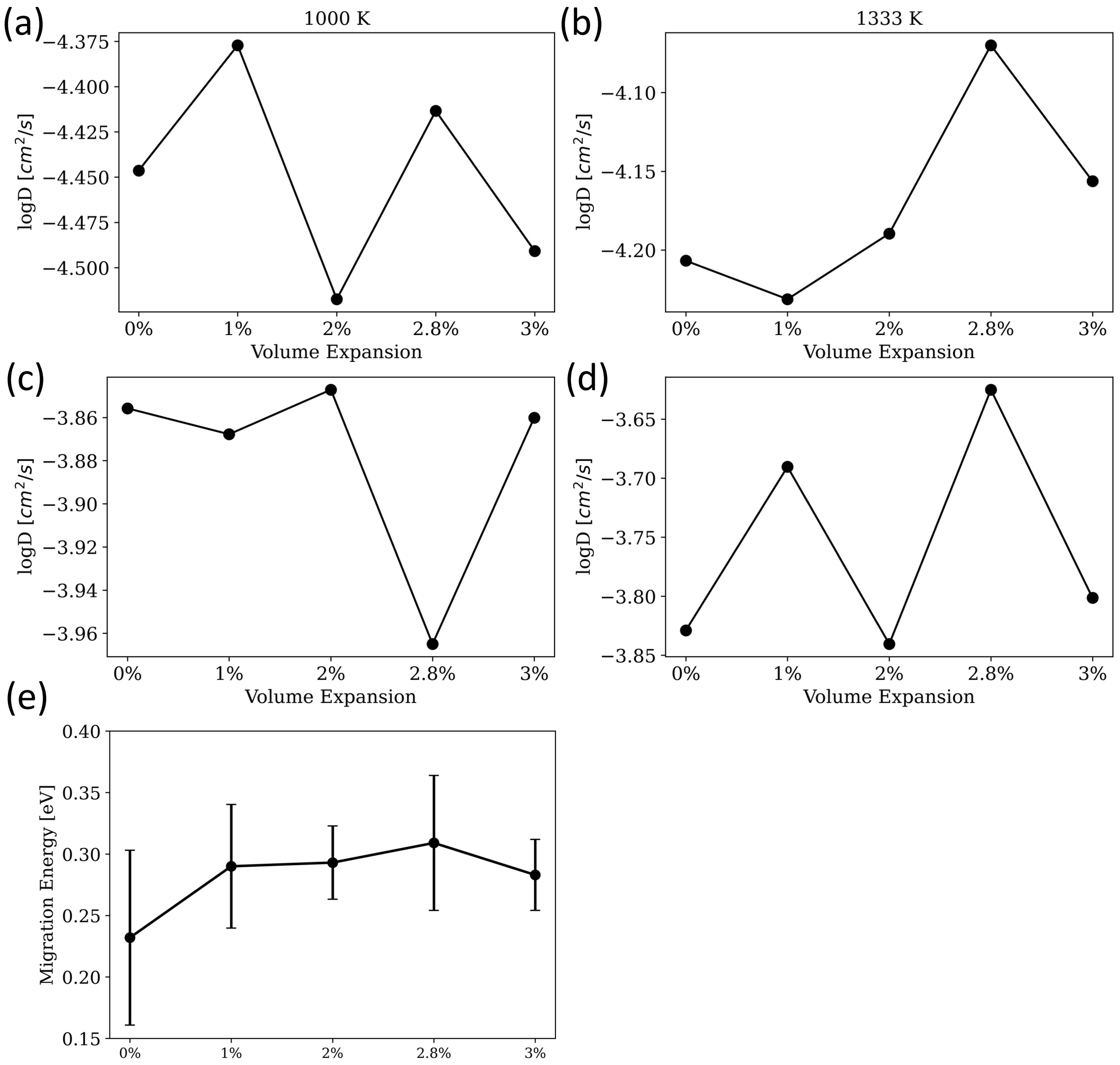

Figure S1 Proton tracer diffusivity of BCFZ at different volume expansions at (a) 1000 K, (b) 1333 K, (c) 1666 K, and (d) 2000 K, and (e) migration energy ($E_m$) as a function of volume expansion.

We examined the effect of volume expansion on the proton tracer diffusivity of BCFZ. The thermal expansion coefficient of BCFZY materials in the temperature range of 400–800 °C has been reported to be $9.24 \times 10^{-6}$ $K^{-1}$. [1] Based on this value, a temperature increase of 1000 K corresponds to an approximately 2.8% volume expansion. To investigate the effect of volume expansion on proton tracer diffusivity in BCFZ, a series of AIMD simulations were performed at volume expansions of 0%, 1%, 2%, 2.8%, and 3%. No clear trend of increasing or decreasing proton tracer diffusivity was observed with increasing volume expansion. In addition, the proton migration energies obtained at different volume expansions were within their respective error bars, indicating no statistically significant differences among the calculated migration energies.

## S2. Proton tracer diffusivity with different oxygen non-stoichiometry in BCFZ

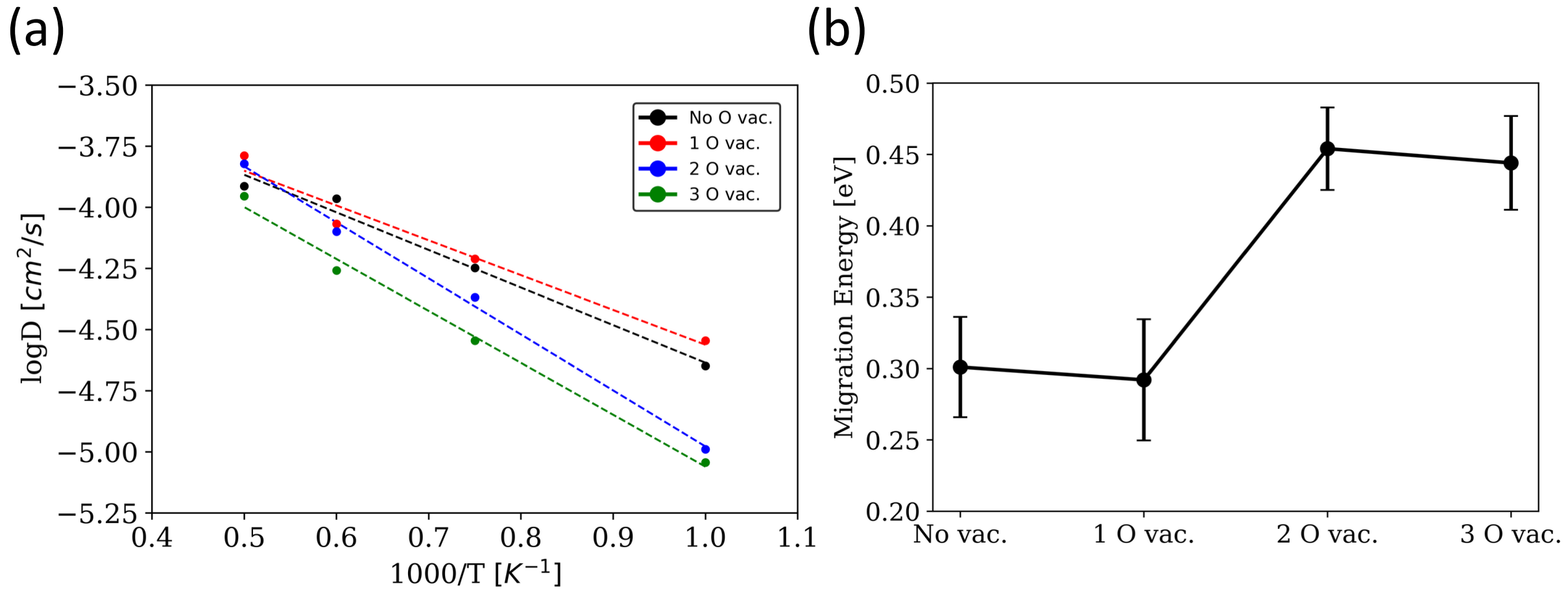

Figure S2. Proton tracer diffusivity (a) and $E_m$ (b) of BCFZ with different numbers of oxygen vacancies

To investigate the influence of oxygen vacancies on proton diffusion in BCFZY systems, a series of AIMD simulations were performed on BCFZ structures with varying oxygen non-stoichiometries (0 to 3 oxygen vacancies). While Y doping is known to introduce oxygen vacancies that can, in principle, impact proton transport, we sought to determine how changes in oxygen vacancy concentration affect proton diffusivity and its associated migration barrier. The calculated proton tracer diffusivities (Figure S2 (a) and migration energies ($E_m$) extracted from Arrhenius fits (Figure S2 (b)) were therefore compared across different oxygen vacancy concentrations. The migration energies were found to be $0.30 \pm 0.04$ eV (no vacancy), $0.29 \pm 0.04$ eV (1 vacancy), $0.45 \pm 0.03$ eV (2 vacancies), and $0.44 \pm 0.03$ eV (3 vacancies). Interestingly, the systems containing two or three oxygen vacancies exhibit higher migration energies than those containing zero or one vacancy, suggesting that increased oxygen vacancy concentration may increase the proton migration barrier. However, further investigation is needed to determine whether this trend represents a systematic effect of oxygen vacancy concentration or arises from other structural and configurational factors.

## S3. B-site configurations for accurate calculation of proton diffusivity

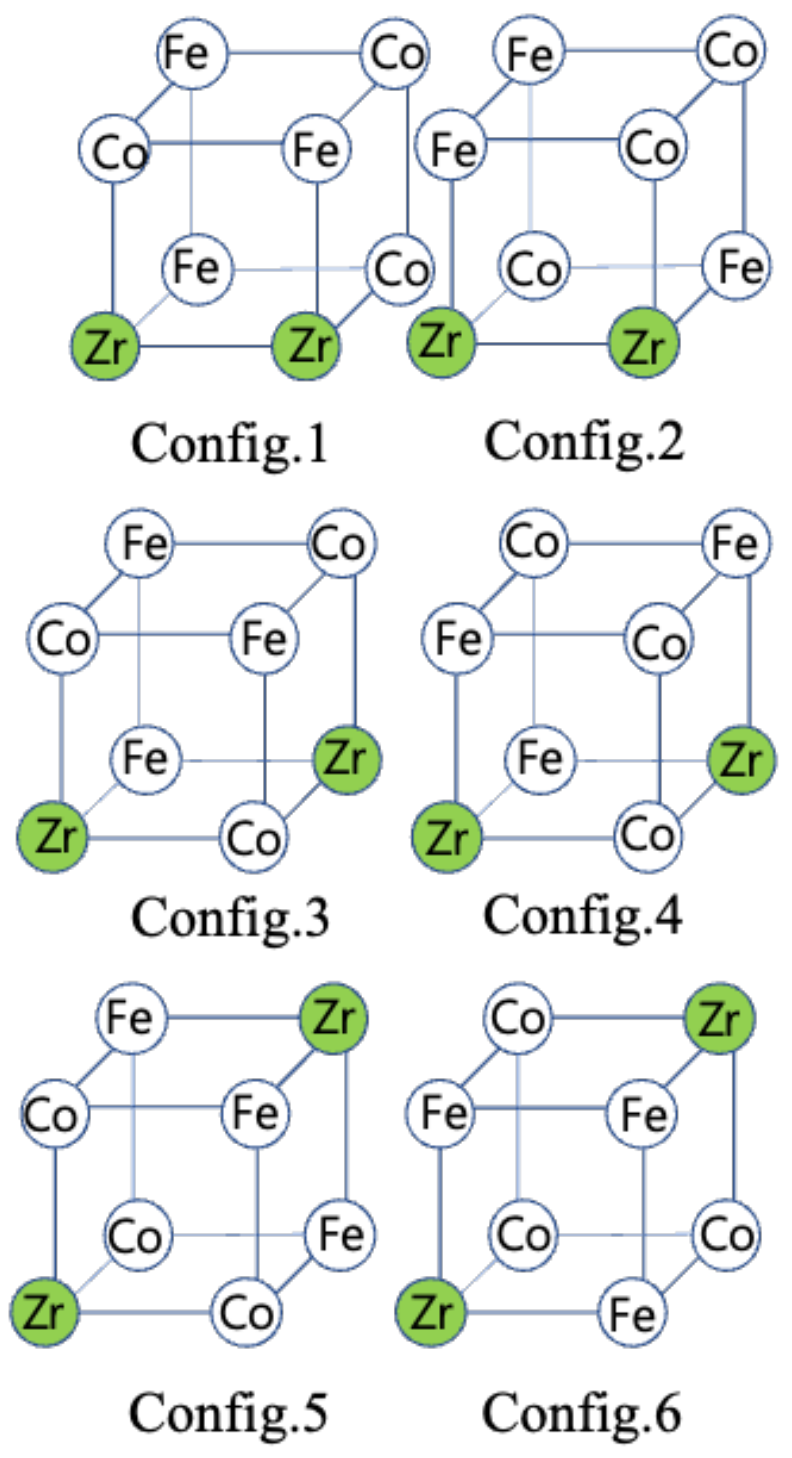

Figure S3. Schematic description of 6 different cation arrangements in BCFZ

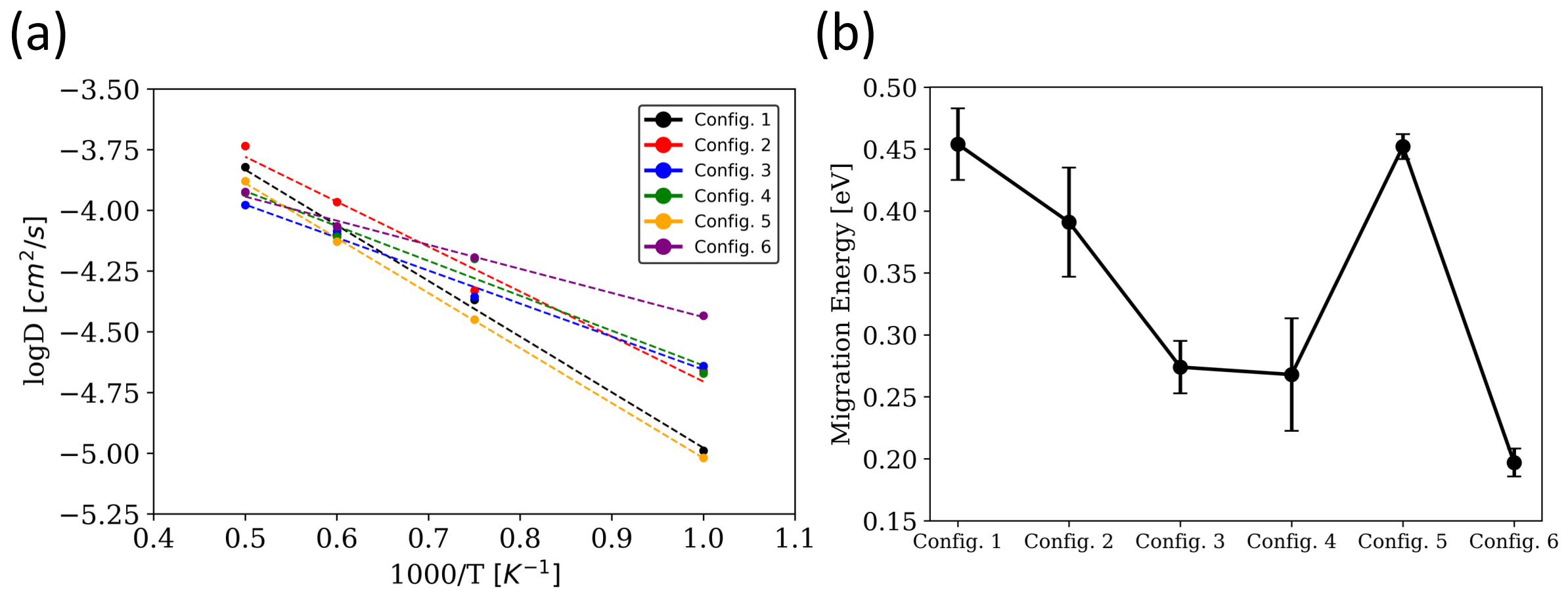

Figure S4. (a) Tracer diffusivity of proton and (b) migration energy for proton diffusion in BCFZ with six different B-site cation configurations.

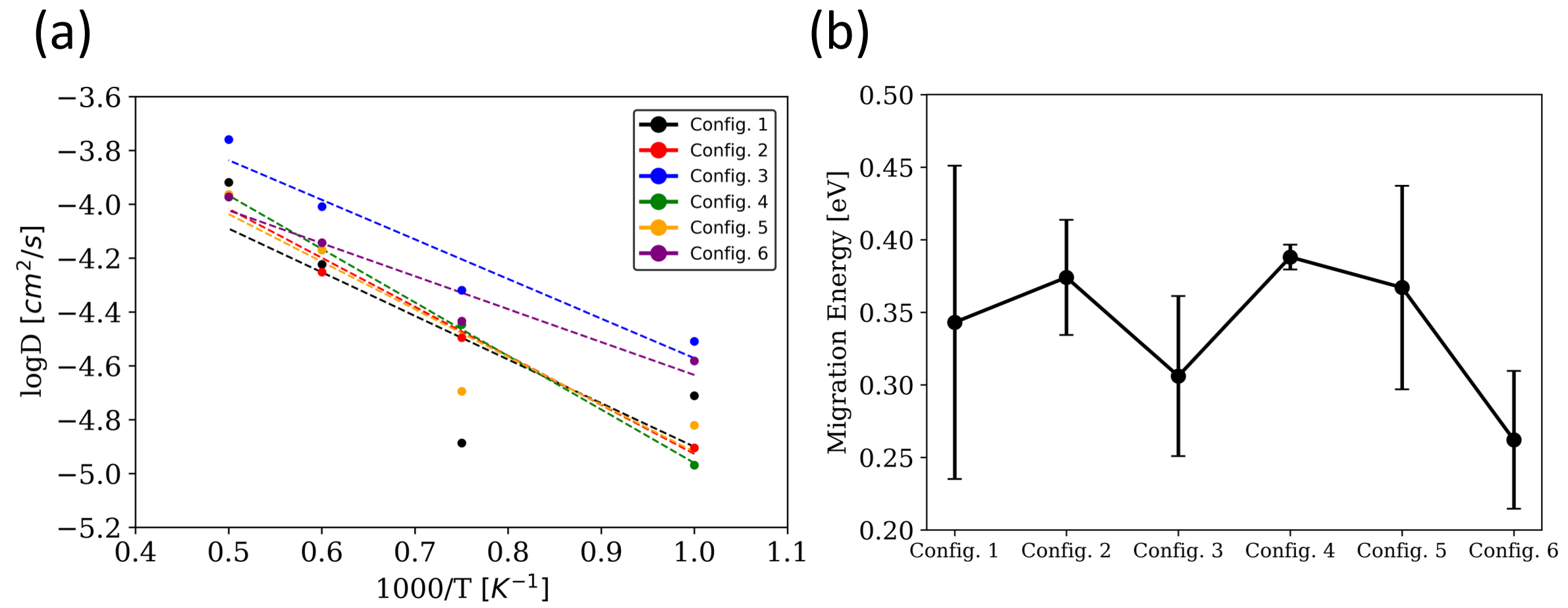

Figure S5. (a) Tracer diffusivity of proton and (b) migration energy for proton diffusion in BCFZY0.1 with six different B-site cation configurations.

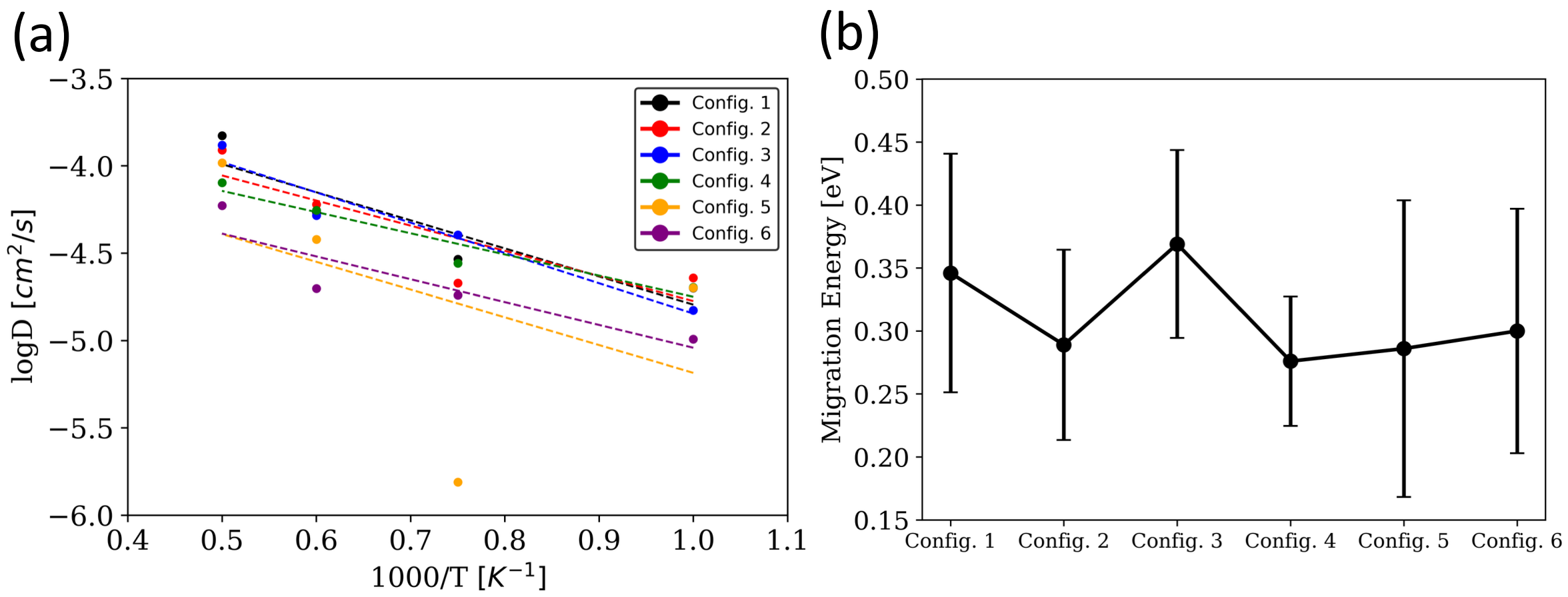

Figure S6. (a) Tracer diffusivity of proton and (b) migration energy for proton diffusion in BCFY with six different B-site cation configurations.

To construct representative atomic configurations of the BCFZY system, six distinct supercell configurations were generated by systematically varying the relative distances between Zr and Y dopants. (Figure S3) The remaining B-site cations (Fe and Co) were randomly distributed across each configuration. To ensure accurate estimation of proton tracer diffusivities in BCFZ (Figure S4), BCFZY (Figure S5), and BCFY (Figure S6), diffusivities were calculated independently for all six configurations and subsequently averaged to reflect the behavior of a disordered solid solution.

## S4. Oxygen vacancy tracer diffusivity in BCFZY materials for applying to hydride diffusion

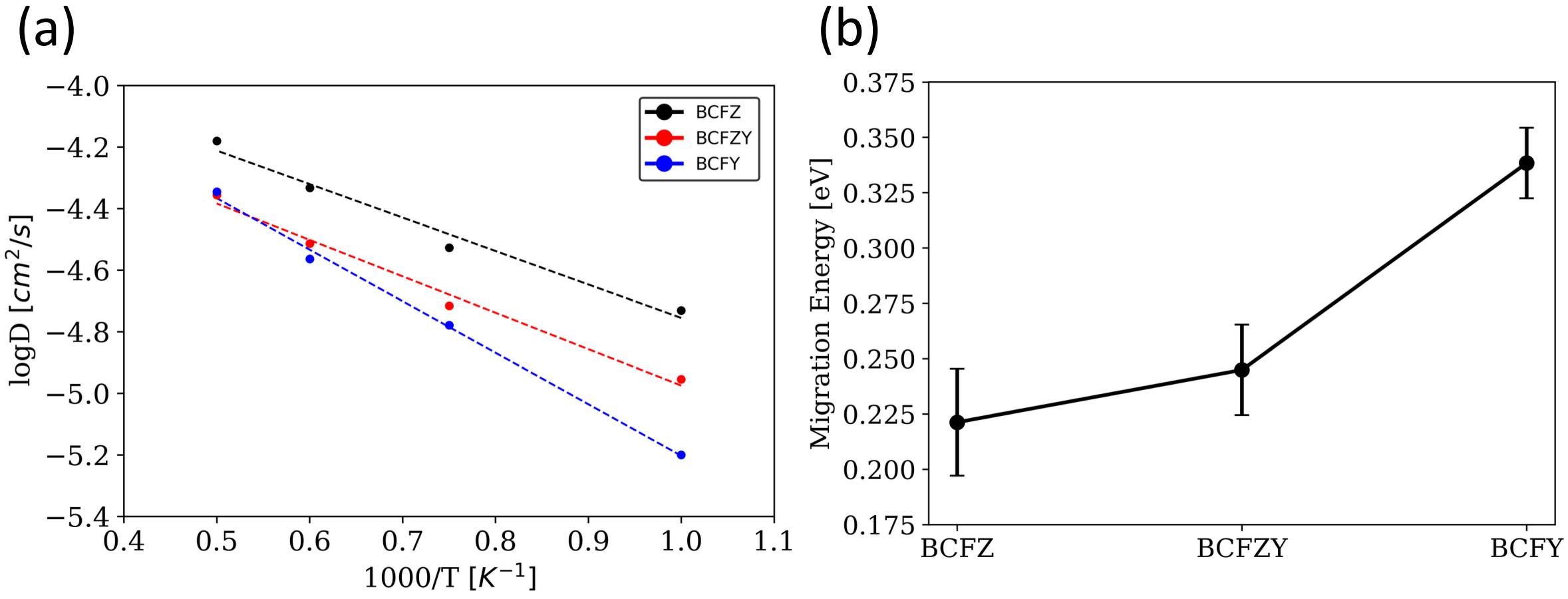

Figure S7. Tracer diffusivity (a) and $E_m$ (b) of oxygen vacancies in BCFZ, BCFZY0.1 and BCFY.

Hydrogen in BCFZY can exist in both hydroxyl ($OH_O^{\cdot}$) and hydride ($H_O^{\cdot}$) forms. To calculate the proton conductivity associated with hydride species using the Nernst–Einstein relation (Equation 19), it is necessary to use the diffusivity of the hydride species. In this work, we assume that the diffusivity of hydride species can be approximated by the oxygen vacancy diffusivity calculated in our previous work [2], since the proton is located at oxygen lattice sites and can therefore be assumed to follow the oxygen vacancy. Figure S7 presents the tracer diffusivity and migration energy of oxygen vacancies in BCFZ, BCFZY0.1, and BCFY. The results suggest that the migration energy increases with increasing Y doping in the BCFZY system.

## S5. System size effect on proton diffusion

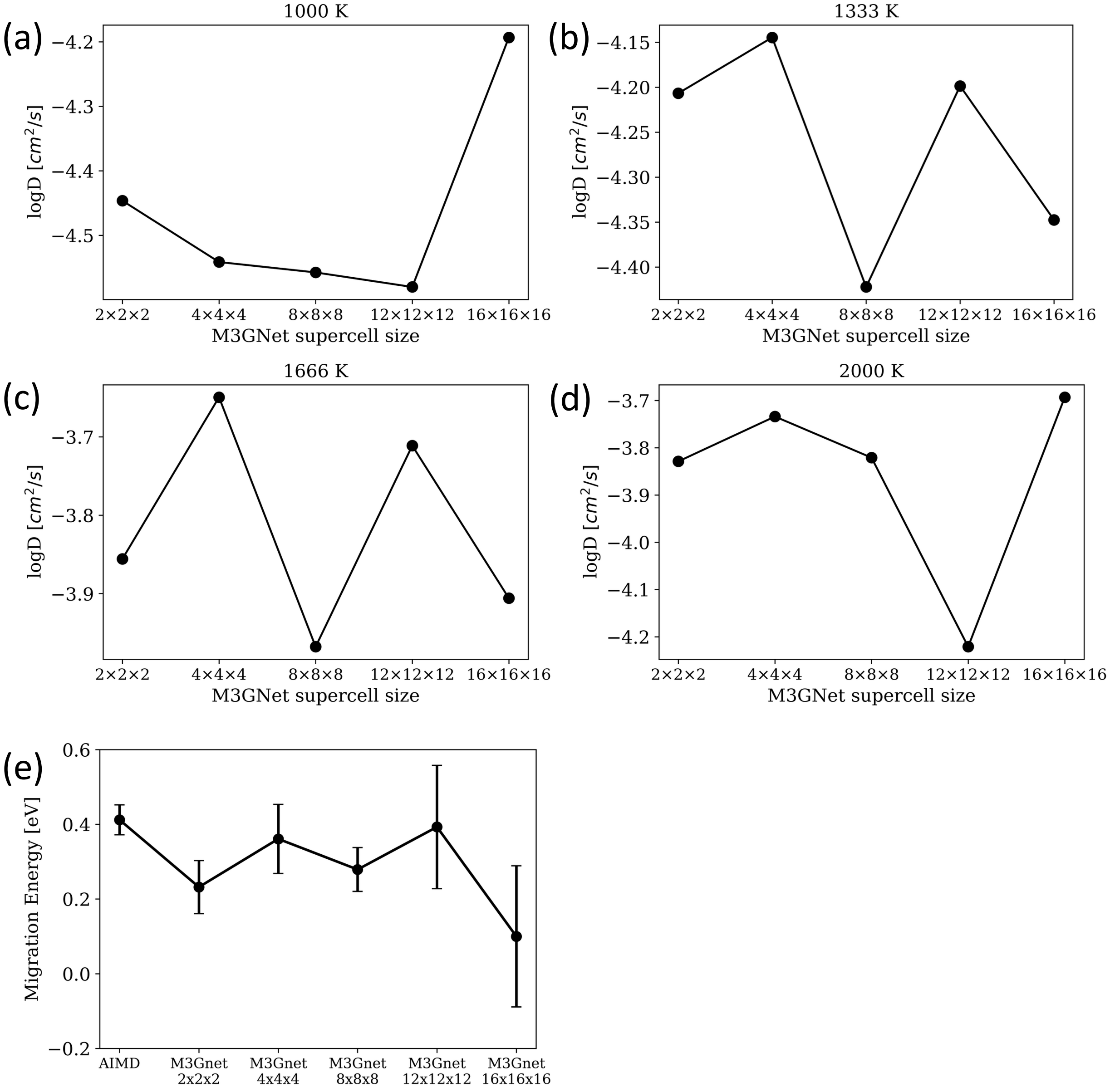

Figure S8. Proton tracer diffusivity of BCFZ at different system sizes at (a) 1000 K, (b) 1333 K, (c) 1666 K, and (d) 2000 K, and (e) migration energy ($E_m$) as a function of system size.

To evaluate the influence of system size on proton diffusion, molecular dynamics simulations were performed using Universal Interatomic Potentials (M3GNet) across multiple supercell sizes. The results indicate that cell size does not exert a significant effect on proton diffusivity, suggesting that finite-size effects are minimal for the time and length scales explored. However, it should be noted that the potential energy surface (PES) softening inherent to M3GNet may lead to an underestimation of migration barriers, which could affect the absolute accuracy of predicted activation energies. [3] Despite this, M3GNet remains a computationally efficient approach for capturing qualitative trends in proton transport across extended length scales.

## S6. Proton conductivity of BCFZY under PCFC operating gas conditions

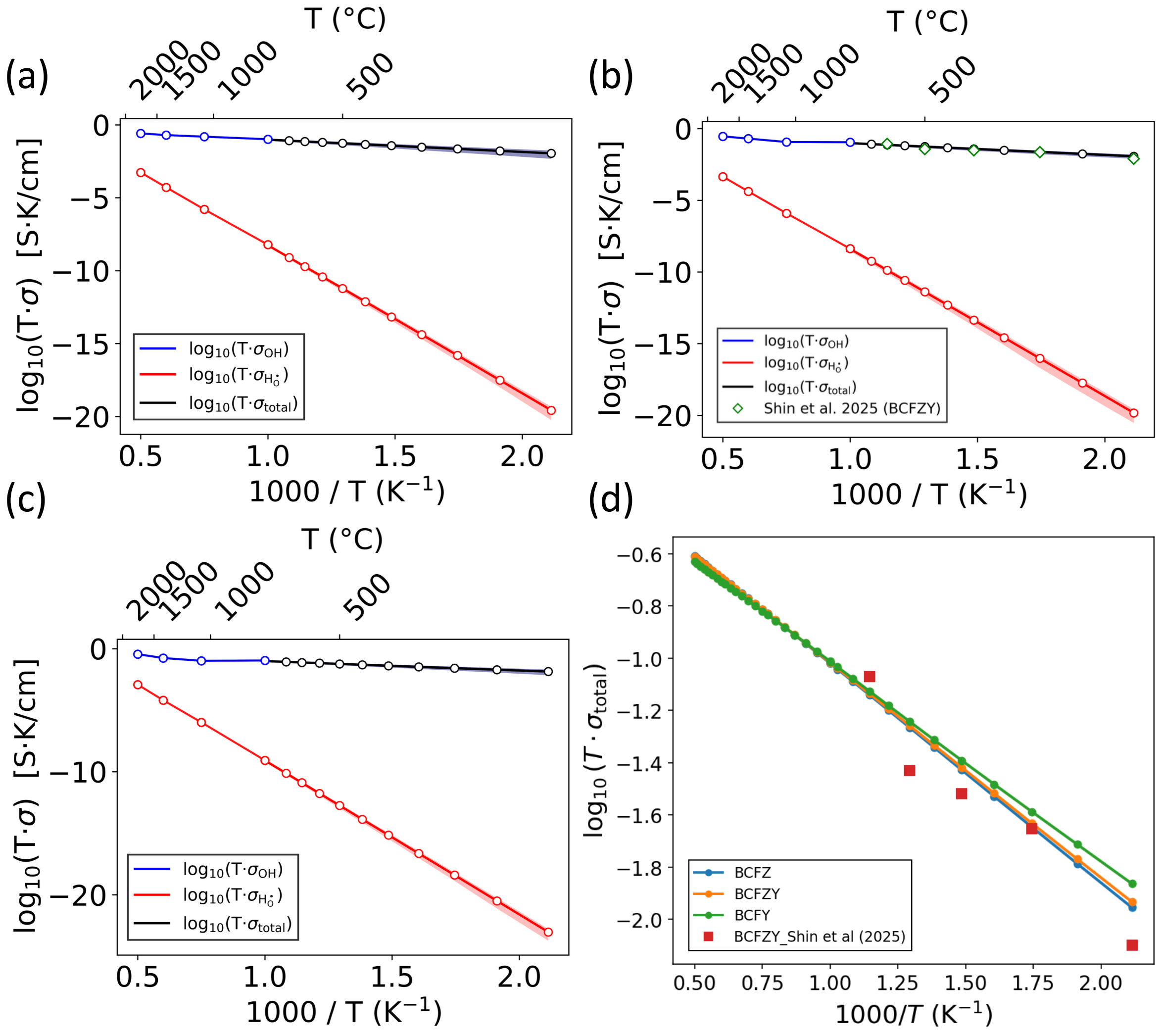

Figure S9. Proton conductivity ($\sigma_H$) of BCFZ, BCFZY0.1, and BCFY under PCFC operating gas conditions for BCFZY0.1. (a) BCFZ, (b) BCFZY0.1, and (c) BCFY. (d) Figure 6 extended to a temperature range up to 2000K.

S7. Proton conductivity of BCFZY under low pO2 conditions.

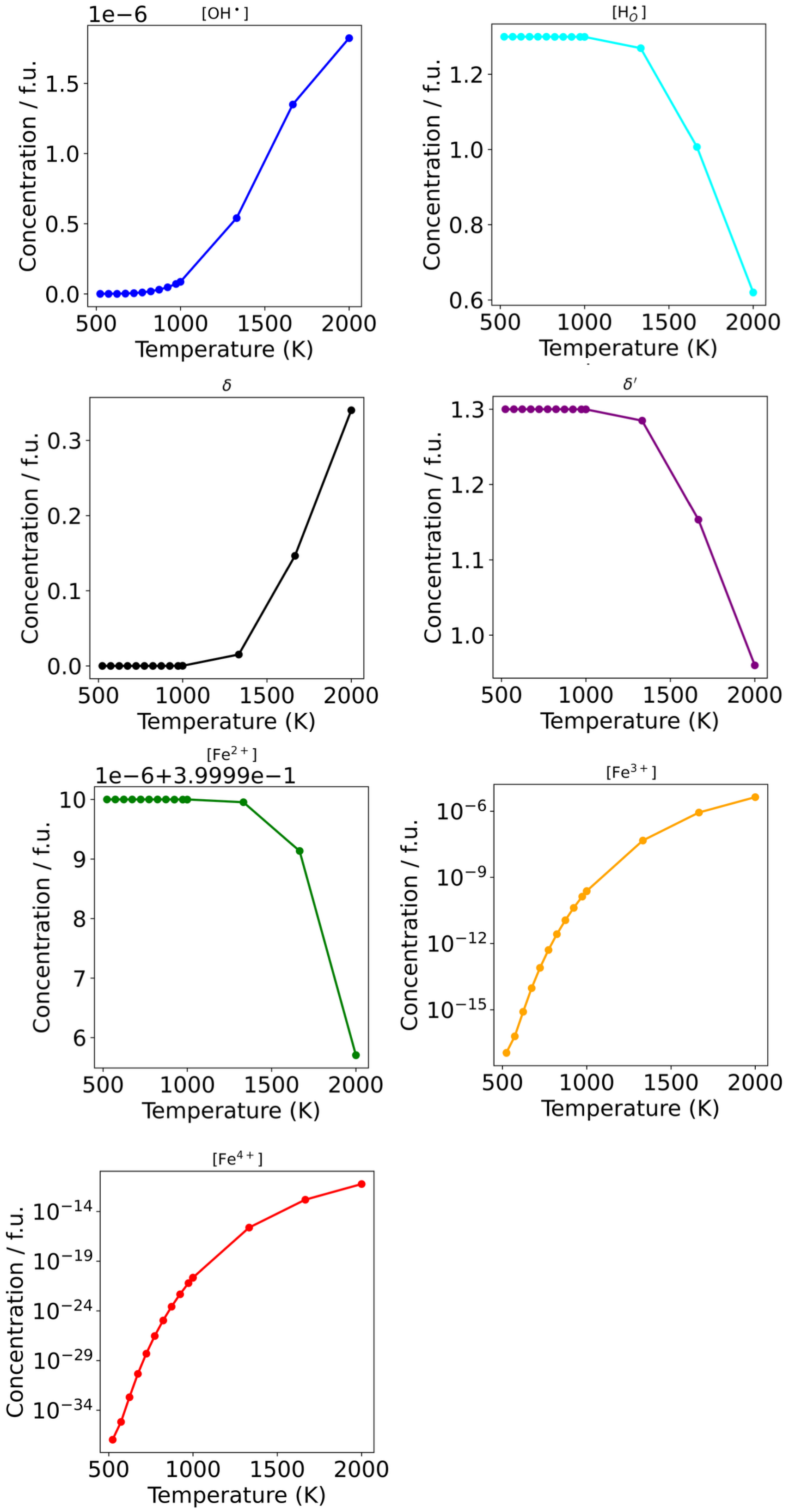

Figure S10. Calculated defect concentrations in BCFZY0.1 under an initial $H_2$ : $O_2$ gas ratio of 0.05 : 0.000001. The calculated species include hydroxyl ($OH_O^{\cdot}$), hydride ($H_O^{\cdot}$), and $Fe^{2+}$, $Fe^{3+}$, and $Fe^{4+}$ oxidation states. Here, δ represents the concentration of empty oxygen vacancies, while δ′ represents the total oxygen deficiency, including hydride species (δ′ = δ +[$H_O^{\cdot}$]).

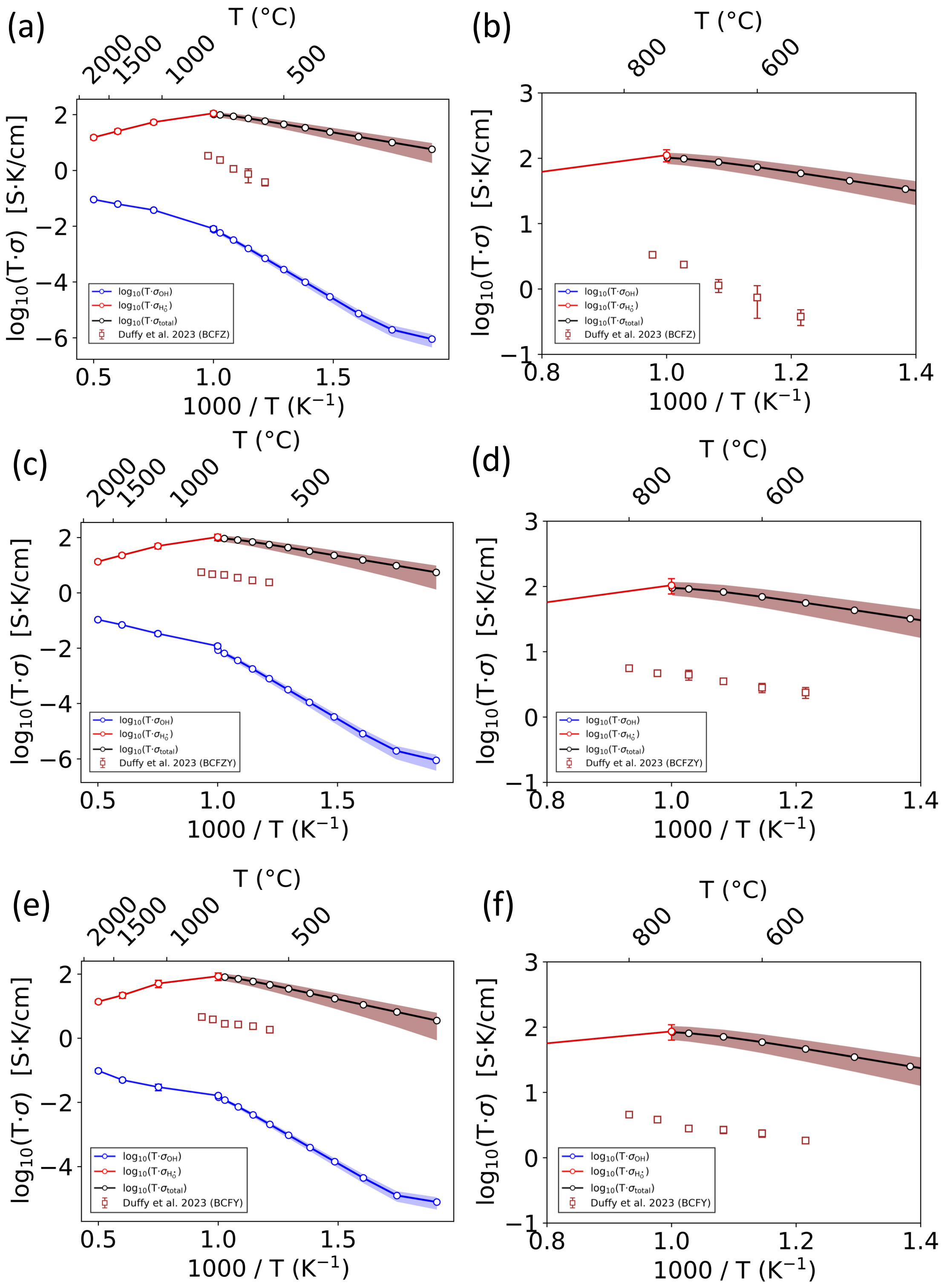

Figure S11. Calculated proton conductivity ($\sigma_H$) of BCFZ, BCFZY0.1, and BCFY under an initial gas ratio of $H_2$ : $O_2$ = 0.05 : 0.001, compared with experimental data. (a) BCFZ, (c) BCFZY0.1, and (e) BCFY over the full temperature range. Panels (b), (d), and (f) show enlarged views of the experimental temperature range for BCFZ, BCFZY0.1, and BCFY, respectively.

Compared with the case of $H_2$ : $O_2$ = 0.05 : 0.000001, the calculated proton conductivity contributed by hydride species ($H_O^{\cdot}$) under $H_2$ : $O_2$ = 0.05 : 0.001 conditions decrease with increasing temperature above approximately 1000 K. This behavior may be explained by the competing effects of hydride concentration and diffusivity. Although hydride diffusivity increases with temperature, the reduction in hydride concentration appears to become sufficiently pronounced above 1000 K to outweigh the increase in diffusivity, leading to a decrease in hydride conductivity. The higher oxygen chemical potential under this condition may also contribute to the reduced stability of hydride species, although further analysis of the defect concentrations is needed to confirm this mechanism.

References

[1] Y. Shin, Y.D. Kim, M. Sanders, S.P. Harvey, M. Walker, R. O'Hayre, Tuning the Co/Fe ratio in BaCoxFe0.8−xZr0.1Y0.1O3−δ, a promising triple ionic and electronic conducting oxide, to boost electrolysis and fuel cell performance, J. Mater. Chem. A Mater. 10 (2022) 24839–24853. https://doi.org/10.1039/d2ta03150g.

[2] C. Kim, R. Jacobs, J.H. Duffy, K.S. Brinkman, H.W. Abernathy, D. Morgan, Effects of Yttrium doping on Oxygen Conductivity in Ba (Fe, Co, Zr, Y) O3-δ Cathode Materials for Proton Ceramic Fuel Cells, Physical Chemistry Chemical Physics 28 (2026) 20805–20815. https://doi.org/10.1039/D5CP04074D.

[3] B. Deng, Y. Choi, P. Zhong, J. Riebesell, S. Anand, Z. Li, K. Jun, K.A. Persson, G. Ceder, Overcoming systematic softening in universal machine learning interatomic potentials by fine-tuning, (2024). https://doi.org/10.1038/s41524-024-01500-6.